\documentclass[12pt]{iopart}
\usepackage{color}
\usepackage{graphicx}
\def\cs2{c_{s}^{2}}

 \def\p{\partial}

 \def\be   {\begin{equation}}   \def\ee   {\end{equation}}
 \def\ba   {\begin{array}}      \def\ea   {\end{array}}
 \def\bea  {\begin{eqnarray}}   \def\eea  {\end{eqnarray}}
 \def\bean {\begin{eqnarray*}}  \def\eean {\end{eqnarray*}}
\begin{document}

\hspace{-30pt} UMN-TH 3205/13 

{\Huge \title{The Trispectrum in the Effective Theory of Inflation with Galilean symmetry}}

\author{ \hspace{-30pt} Nicola Bartolo$^{1,2}$, Emanuela Dimastrogiovanni$^{3}$ and Matteo Fasiello$^{4}$}
\vspace{0.4cm}
\address{\hspace{-30pt}$^1$ Dipartimento di Fisica e Astronomia ``G. Galilei'', Universit\`{a} degli Studi di 
Padova,  \\
via Marzolo 8, I-35131 Padova, Italy $\qquad\qquad\qquad\qquad\qquad\qquad\qquad$} 
\address{\hspace{-30pt}$^2$ INFN, Sezione di Padova, via Marzolo 8, I-35131 Padova, Italy}
\address{\hspace{-30pt}$^3$ School of Physics \& Astronomy, University of Minnesota, Minneapolis, 55455, USA }
\address{\hspace{-30pt}$^4$ Department of Physics, Case Western Reserve University, Cleveland, 44106, USA } 
\vskip 0.5cm

\eads{\mailto{nicola.bartolo@pd.infn.it}, \mailto{emanuela1573@gmail.com},  \mailto{mrf65@case.edu} }

\begin{abstract}
We calculate the trispectrum of curvature perturbations for a model of inflation endowed with Galilean symmetry at the level of the fluctuations around an FRW background. Such a model has been shown to posses desirable properties such as unitarity (up to a certain scale) and non-renormalization of the leading operators, all of which point towards the reasonable assumption that a full theory whose fluctuations reproduce the one here might exist  as well as be stable and predictive. \\
The cubic curvature fluctuations of this model produce quite distinct signatures at the level of the bispectrum. Our analysis shows how this holds true at higher order in perturbations.  We provide a detailed study of the trispectrum shape-functions in different configurations and a comparison with existent literature. Most notably, predictions markedly differ from their $P(X,\phi)$ counterpart in the so called \textit{equilateral} trispectrum configuration. \\
The zoo of inflationary models characterized by somewhat distinctive predictions for higher order correlators is already quite populated; what makes this model more compelling resides in the above mentioned stability properties.

\end{abstract}
\newpage

\tableofcontents

\section{Introduction}
The inflationary paradigm \cite{Lyth:1998xn} stands at the core of modern cosmology. It set out to solve several cosmological puzzles which pre-dated it and delivered further detailed predictions verified to be in close agreement with observations \cite{Bennett:2012fp,Hinshaw:2012fq,Ade:2013rta,Ade:2013ydc}. The simplest models of single-field slow-roll inflation are, as of today, in agreement with data. On the other hand, it is natural to ask whether it is possible to find or embed an inflationary mechanism within UV complete theories such as string theory. One such a realization (DBI inflation) was put forward in~\cite{DBI,Chen:2004gc,Chen:2005ad}, and has since been studied in great details (for other realizations see~\cite{Kachru:2003sx},~\cite{Silverstein:2008sg,McAllister:2008hb,Flauger:2009ab,Berg:2009tg} and \cite{Cline:2006hu,Kallosh:2007ig,Burgess:2011fa} for some reviews). However, the regime in which the DBI model can be trusted seems perilously close to what data might soon exclude.\\

 Stopping a bit short of this last, quite ambitious, requirement, one could also aim at a model of inflation which is \textit{unitary}, \textit{stable} under quantum corrections and \textit{predictive}. In \cite{Andrew1} the authors propose 
 an inflationary set up in this direction, where the requirement of the so called \textit{Galileon} symmetry ($\phi \rightarrow \phi + c + b_{\mu} x^{\mu}$, see~\cite{Nicolis:2008in} ) grants all of the above properties: \\

\noindent - in flat space the $n$ (where the index $n$ counts space-time dimensions) Galilean terms are guaranteed to have second order equations of motion \cite{Nicolis:2008in} \\

\noindent - the non-renormalization theorem for Galileon models \cite{Nicolis:2008in} warrants that (at least in flate space) the coefficients which one writes down at first will not be largely modified by renormalization.\\

\noindent Crucially, these desirable properties are shown to approximately (up to $ \Lambda/M_{Pl}$ corrections, where $\Lambda$ is the scale of the underlying theory) hold for the model in \cite{Andrew1} also in the more realistic scenario where one deals with curved spacetime and couples to gravity. There exists an interesting energy regime where the Galileon interactions are important and the model is quite different from canonical slow-roll inflation. In this same regime both the metric fluctuations and terms with non-minimal coupling to gravity can be neglected and calculating higher order curvature correlators is greatly simplified.

 The bispectrum of this model has been explored in \cite{Andrew1}: the non-linearity parameter $f_{NL}$ generated can be detectable but the shape-functions peak in the equilateral configuration and thus one is not able to remove the large degeneracy with most other inflationary models at the level of the bispectrum. This, of course, leaves the appeal of the model just described untouched.
One should then go look for another quantity at (possibly) observational  reach, the trispectrum being the natural choice. This is precisely what we do in a paper \cite{Fred} which is complementary to the work presented here.

On the other hand, if one is prepared to give away a little bit in terms of the cherished properties described above (this statement will be made more quantitative below), then there exist a model \cite{Creminelli1} which already at the bispectrum level generates  interesting and quite distinct shape signatures. Combining the analysis of \cite{Creminelli1} with the study of the four point function, the trispectrum, one can really hope to fully characterize the predictions of this \textit{generalized Galileon} model and disentangle it from a host of inflationary models with different, somewhat more canonical, signatures.\\

\noindent Briefly, in going from the approach of~\cite{Andrew1} (see also \cite{Fasiello:2013dla}) to the one of~\cite{Creminelli1} one: 

\noindent - has a smaller scale $\Lambda$ up to which the effective theory can be trusted\\
- with some assumptions (reasonable in light of \cite{Andrew1} and other works), preserves the non-renormalization properties of the Galileon terms \\
- has a richer and more distinct set of predictions for non-Gaussianities\\

\noindent What has been proven concerning the background (and therefore the full theory) in \cite{Andrew1} is more or less implicitly assumed at the onset of \cite{Creminelli1} because the latter is an effective field theory of fluctuations around a background rather than a effective field theory as a whole. The focus is on fluctuations and their properties.

Following \cite{Creminelli1} then, we study the four-point function of curvature perturbations in detail, 
we calculate the corresponding amplitude and plot the shape-functions of each interaction term in different configurations so as to ease the comparison with the existent literature on the subject and, at least at the level of shape-functions, remove as much  degeneracy as possible.

Our analysis nicely fits within the realm of the effective field theory approach to inflation \cite{Creminelli:2006xe,Cheung,Bartolo:2010bj,Bartolo:2010di,Senatore:2010wk,Bartolo:2010im, Baumann:2011su,Fasiello:2011fj,Behbahani:2011it,LopezNacir:2012rm}. See \cite{Weinberg:2008hq} for a different approach.

On the other hand, the model we study  is not quite as generic as it could have been in the spirit of \cite{Cheung}: we are now dealing with a theory which is unitary (up to some scale) and has been further endowed with a symmetry which selects interaction operators and protects them from quantum corrections. This subset of theories is more stable and more predictive in that the symmetry shields us from large renormalization of the coefficients of interaction terms and keeps the number of leading interaction operators to a finite, small number.\\

The paper is organized as follows.  In Sec.~2 we introduce the action for the model and expand it up to fourth order in fluctuations, we also perform a field redefinition which returns the action in a form that makes it suitable for employment  within the Hamiltonian formulation of the Schwinger-Keldysh formalism. In Sec.~3, we compute the trispectrum of curvature fluctuations, study its amplitude and shape functions so as to draw a comparison with some of the results already present in the literature for other inflationary models, especially DBI inflation (for studies on primordial trispectrum in DBI, and more general $P(X,\phi)$ models, and other inflation models, see, e.g., \cite{Seery:2006vu,Huang:2006eha,Seery:2006js,Byrnes:2006vq,Arroja:2008ga,Seery:2008ax,Mizuno:2009cv,Gao:2009gd,Chen:2009bc,Arroja:2009pd,Mizuno:2009mv}). In Sec.~4 we conclude, and offer some further comments. In Appendix~A  we present the details of our field redefinition while in Appendix~B we review the main steps that characterize the IN-IN formalism computation of the trispectrum and we list the final analytical results for the shape functions. In the following greek indices run from $0$ to $3$, latin indices are spatial only, running from $1$ to $3$.

\section{The Action to fourth order in perturbations}


\subsection{Building Blocks of the Action}
We are interested in a Lagrangian for scalar perturbations $\pi$ around an FRW background whose fluctuations satisfy the Galilean symmetry. The model can be written down by listing all possible Galilean invariant terms which non-linearly realize Lorentz symmetry. 

Alternatively, one can write down Lorentz invariant contributions in terms of $\psi=t+\pi(\vec x,t)$ where $\psi$ linearly realizes the symmetry \cite{Creminelli1}. We shall proceed according to the latter prescription. What we are after is the number of independent operators generated by products of traces of the form:
\bea
\left[\Psi^{n_1} \right]^{k_1} \times \left[\Psi^{n_2} \right]^{k_2}\times ... \label{build}
\eea
\noindent where each $\left[... \right] $ is to be understood as a trace of whatever product is inside the brackets, $\Psi=\nabla_{\mu}\nabla_{\nu} \psi$, and indices are raised with the metric $g_{\mu\nu}$ . It can be shown that, once all the terms of the form $\left[\Psi^{n} \right]$ with $n \leq 4$ are known, all the rest in Eq.~(\ref{build}) can be expressed as linear combinations of products of these building blocks \cite{Creminelli1}.   \\
With fluctuations in mind, one removes the background value  $c_n$ from each trace below:
\bea 
\fl a_1: \left[\Psi^1 \right]-c_1= \nabla_{\mu}\nabla^{\mu}\psi -c_1=\nabla_{\mu}\nabla^{\mu}\pi  \nonumber \\
\fl a_2: \left[\Psi^2 \right]-c_2=  \nabla_{\mu\nu}^{}\pi\nabla_{}^{\nu\mu}\pi-2 H \nabla_{i}\nabla^i \pi  \nonumber\\
\fl a_3: \left[\Psi^3 \right]-c_3=  \nabla_{\mu \nu}^{} \pi \nabla_{}^{\nu \rho}\pi \nabla_{\rho}^{\mu}\pi -3 H \nabla_{\mu i}^{} \pi \nabla_{}^{i \mu}\pi  +3 H^{2} \nabla_{i}\nabla^i \pi    \nonumber\\
\fl a_4: \left[\Psi^4 \right]-c_4=...-4 H \nabla_{\mu i}^{} \pi \nabla_{}^{i \nu}\pi \nabla_{\nu}^{\mu}\pi +4 H^{2} \nabla_{\mu i}^{} \pi \nabla_{}^{i \mu}\pi +2 H^{2}  \nabla_{ij}^{} \pi \nabla_{}^{ji}\pi -4 H^{3} \nabla_{i}\nabla^i \pi \nonumber\\
\label{an}
\eea
\noindent where the expressions of the type $\nabla_{\mu_1 \mu_2}\pi \,\, ; \,\, \nabla_{\mu_1}^{ \mu_2}\pi $  are meant as shortcuts for $\nabla_{\mu_1}\nabla_{\mu_2}\pi \,\, ; \,\, \nabla_{\mu_1}\nabla^{\mu_2}\pi$ (greek indices run from $0$ to $3$, latin indices are spatial only) . It is easy to check that Eq.~(\ref{an}) represents a basis which spans the entire space (up to order $\pi^3$) of the $\left[\Psi^n \right]-c_n$ operators for any $n\geq 4$ . 

\noindent In what follows we do not consider contributions coming from a single trace e.g. $\left[\Psi^{\,n} \right]^{1}$; the reason is that these will account \cite{Creminelli1} for the so-called \textit{minimal covariant Galileon terms} \cite{Deffayet1,Deffayet:2009mn,Deffayet:2010zh} and those, their expression being known, can be put in at a later stage. More explicitly, the \textit{single trace} operators  can be written as sums of multiple trace operators and minimal Galileon terms. The former are already all accounted for in Eq.~(\ref{an}) while the latter can be shown to be subdominant whenever generic $(\nabla^2 \psi)^n$ operators are switched on, as is the case here~\cite{Creminelli1}.\\
\noindent With Eq.~(\ref{an}) handy, one sets about to write down all the leading independent interactions at third and fourth order in perturbations. Particular care must be exerted so as to have that, at each perturbative order, the coefficient multiplying  a given interaction term does not appear already at lower perturbative order. Such an occurrence is not problematic by itself but, as is known \cite{Creminelli1}, the third and fourth order contribution of a piece like $M^4\left(\dot \pi^2 +\dot \pi^3+ \dot \pi^4 \right)$ would be suppressed by  a scale $\Lambda^{\prime}$ larger than the one $\Lambda<\Lambda^{\prime}$ that suppresses the quadratic fluctuations. The resulting three and four-point functions would therefore be negligible with respect to the contribution coming from interactions suppressed by $\Lambda$. 

\noindent Indeed, what is natural to assume at this stage is that:\\
\textit{Every independent interaction term at each given perturbative order is suppressed by a common scale $\Lambda$} . This is possible if each coefficient which corresponds to each independent interaction appears at given perturbative order and nowhere else \footnote{ That is, appears as multiplying a leading operator.}. 

This requirement is relatively easy to implement. Using the building blocks in Eq.~(\ref{an}), what one wants is independent combinations of product of multiple trace operators whose first contribution starts at a given perturbative order
\bea
\mathcal{O}_3= \sum_{\vec x} c_{\tilde M}(x_1,x_2,x_3)\times  a_{1}^{x_1} a_{2}^{x_2} a_{3}^{x_3}a_{4}^{x_4}= \tilde M \times \mathcal{O}(\pi^3)+...\\
\mathcal{O}_4= \sum_{\vec x} c_{\bar M}(x_1,x_2,x_3,x_4)\times  a_{1}^{x_1} a_{2}^{x_2} a_{3}^{x_3} a_{4}^{x_4}= \bar M \times \mathcal{O}(\pi^4)+... , 
\eea
where the $c$'s are constant (to first approximation in generalized slow-roll) coefficients. As an example, consider:
\bea
\fl\quad  a_2^3\sim \tilde{M}_4 (\nabla_{i}\nabla^{i}\pi)^3 ; \quad a_1^4\sim \bar{M}_5 (\nabla_{\mu}\nabla^{\mu}\pi)^4; \quad \left(a_4+ \frac{4}{3} H a_3 \right)^2 \sim \bar{M}_1 \left(\nabla_{i j}\pi \nabla^{i j} \pi\right)^2  .
\eea
The coefficients $\tilde{M}_k, \bar{M}_n$ are all different and appear only once in our total Lagrangian. Applying this algorithm to the fullest, one obtains the interactions Lagrangians below.

\subsection{Interaction Lagrangian}
\noindent We write below the expression for the $\mathcal{S}_3, \mathcal{S}_4$ action in $\pi$:
\bea
\label{ac3}
\fl \mathcal{S}_3=\int dt d^{3}x \sqrt{-g} \{   \tilde{M}_{1}(\nabla_{\mu}^{2}\pi)^3+\tilde{M}_{2}(\nabla_{\mu}^{2}\pi)^2(\nabla_{i}^{2}\pi) +\tilde{M}_{3}(\nabla_{\mu}^{2}\pi)(\nabla_{i}^{2}\pi)^2+ \tilde{M}_{4}(\nabla_{i}^{2}\pi)^3 \nonumber \\  
\fl \qquad \qquad \qquad\,\,\,+\tilde{M}^{}_{5} \big[ (\nabla_{\mu \nu} \pi)^2 -2 (\nabla_{ \rho i} \pi)^2  \big](\nabla_{\sigma}^{2}\pi)+\tilde{M}^{}_{6} \big[(\nabla_{\mu \nu} \pi)^2 -2 (\nabla_{ \rho i} \pi)^2  \big](\nabla_{j}^{2}\pi) \nonumber \\
\fl \qquad \qquad \qquad\,\,\, +\tilde{M}_{7}(\nabla_{ij}\pi)^2 (\nabla_{\mu}^{2}\pi)+ \tilde{M}_{8}(\nabla_{ij}\pi)^2 (\nabla_{k}^{2}\pi)  \}
\eea
\bea
\label{ac4}
\fl \mathcal{S}_4=\int dt d^{3}x\sqrt{-g} \{ \bar{M}^{}_{1} \big[(\nabla_{i j}  \pi)^2\big]^2+\bar{M}^{}_{2}(\nabla_{i j} \pi)^2(\nabla_{\mu}^{2}\pi)^2+\bar{M}^{}_{3}(\nabla_{\mu}^{2}\pi)^4+\bar{M}^{}_{4} (\nabla_{i}^{2}\pi)^4  \nonumber \\
\fl \qquad \qquad \qquad\,\,\, +\bar{M}^{}_{5} (\nabla_{i}^{2}\pi)^3(\nabla_{\mu}^{2}\pi)+\bar{M}^{}_{6} (\nabla_{i}^{2}\pi)^2(\nabla_{\mu}^{2}\pi)^2+\bar{M}^{}_{7} (\nabla_{i}^{2}\pi)(\nabla_{\mu}^{2}\pi)^3 \nonumber \\
\fl \qquad \qquad \qquad\,\,\, +\bar{M}^{}_{8} (\nabla_{i}^{2}\pi)^2  (\nabla_{j k} \pi)^2  +\bar{M}^{}_{9} (\nabla_{i}^{2}\pi) (\nabla_{j k} \pi)^2 (\nabla_{\mu}^{2}\pi) \nonumber \\
\fl \qquad \qquad \qquad\,\,\, +\bar{M}^{}_{10} \Big[ \left((\nabla_{\mu \nu} \pi)^2\right)^2 +4\left((\nabla_{\mu i} \pi)^2\right)^2 -4 (\nabla_{\nu \rho} \pi)^2 (\nabla_{\mu i} \pi)^2  \Big]  \nonumber \\
\fl \qquad \qquad \qquad\,\,\,  +\bar{M}^{}_{11} \big[(\nabla_{\mu \nu} \pi)^2 -2 (\nabla_{ \rho i} \pi)^2  \big](\nabla_{\sigma}^{2}\pi)^2 +\bar{M}^{}_{12} \big[ (\nabla_{\mu \nu} \pi)^2 -2 (\nabla_{ \rho i} \pi)^2  \big](\nabla_{j}^{2}\pi)^2 \nonumber \\
\fl \quad \quad\,\,\, +\bar{M}^{}_{13} \big[(\nabla_{\mu \nu} \pi)^2 -2 (\nabla_{ \rho i} \pi)^2  \big](\nabla_{j}^{2}\pi)(\nabla_{\sigma}^{2}\pi)+\bar{M}^{}_{14} \big[ (\nabla_{\mu \nu} \pi)^2 -2 (\nabla_{ \rho i} \pi)^2  \big](\nabla_{jk}\pi)^2\}  \nonumber\\
\eea
The second order Lagrangian is
\bea
\mathcal{S}_2=\int dt d^{3}x \sqrt{-g} M_{P}^{2}\dot{H}\left( \p_{\mu}\pi\p^{\mu}\pi\right) .
\eea
If one were to write down the $\pi$ equations of motion for, say, $\mathcal{S}_3$ above, those would surely include terms like  third time derivatives of the scalar $ d^3 \pi/dt^3$; that is because the action above comprises higher time derivative terms. If taken literally,  this fact leads in turn to the excitation of unstable modes in the $\pi$ dynamics and renders the Hamiltonian of the system unstable \cite{Ostro}. In perturbation theory, it is however possible to treat this instability by essentially connecting the ``higher derivative expansion''  with the perturbative expansion in the field \cite{JLM}.  In the case at hand the latter procedure corresponds to performing an order-by-order (in perturbations) \textit{field redefinition}.  If one stops at third order in fluctuations, it can further be shown that a field redefinition is equivalent (up to $\mathcal{O}(\pi^4)$ corrections) to the repeated use of the $\mathcal{S}_2$-derived  equation of motion to get rid of the time derivatives in excess. This is where we part ways with the analysis in \cite{Creminelli1}: there the bispectrum was the target and the use of the e.o.m.'s was entirely justified to this aim, here we also want to calculate the trispectrum  so we have to undergo (as also recognized in \cite{Creminelli1})  a full fledged field redefinition:  the $\mathcal{O}(\pi^4)$ corrections do matter for our purposes.\\

\noindent Let us redefine the field $\pi$ as follows
\bea
\label{frr}
\pi\rightarrow\pi_{(0)}+\pi_{(1)}+\pi_{(2)}+...,
\eea

\noindent where $\pi_{(1)}\sim\mathcal{O}\left(\pi_{(0)}^{2}\right)$, $\pi_{(2)}\sim\mathcal{O}\left(\pi_{(0)}^3\right)$. We are computing the action up to fourth order in the field fluctuations, therefore we do not need to go any further than $\pi_{(2)}$. If we plug the expansion in Eq.~(\ref{frr}) in the action up to fourth order in $\pi$, we have

\bea
\mathcal{S}_{2}[\pi]\rightarrow \mathcal{S}_{2}^{[2]}[\pi_{(0)}^{2}]+\mathcal{S}_{3}^{[2]}[\pi_{(0)}\pi_{(1)}]+\mathcal{S}_{4}^{[2]}[\pi_{(1)}^{2}]+\mathcal{S}_{4}^{[2]}[\pi_{(0)}\pi_{(2)}]+...  ,\\
\mathcal{S}_{3}[\pi]\rightarrow \mathcal{S}_{3}^{[3]}[\pi_{(0)}^{3}]+\mathcal{S}_{4}^{[3]}[\pi_{(0)}^{2}\pi_{(1)}]+...  ,\\
\mathcal{S}_{4}[\pi]\rightarrow \mathcal{S}_{4}^{[4]}[\pi_{(0)}^{4}]+...,
\eea

\noindent where, on the right-hand side, the lower index on $\mathcal{S}$ indicates the perturbative order whereas the upper index reminds us from which order in $\mathcal{S}[\pi]$ that specific contribution arises. The ``...'' indicates contributions that are beyond fourth order and which we do not need for our computation of the trispectrum. The third and fourth order action in the newly redefined field $\pi_{(0)}$ are then 

\bea
\label{s2}
\mathcal{S}_{2}[\pi_{(0)}]=\mathcal{S}_{2}^{[2]}[\pi_{(0)}^{2}],
\eea
\bea
\label{s3}
\mathcal{S}_{3}[\pi_{(0)}]=\mathcal{S}_{3}^{[3]}[\pi_{(0)}^{3}]+\mathcal{S}_{3}^{[2]}[\pi_{(0)}\pi_{(1)}],
\eea
\bea
\label{s4}
\mathcal{S}_{4}[\pi_{(0)}]=\mathcal{S}_{4}^{[4]}[\pi_{(0)}^{4}]+\mathcal{S}_{4}^{[2]}[\pi_{(1)}^{2}]+\mathcal{S}_{4}^{[2]}[\pi_{(0)}\pi_{(2)}]+\mathcal{S}_{4}^{[3]}[\pi_{(0)}^{2}\pi_{(1)}].
\eea

\noindent We will now determine $\pi_{(1)}$ and $\pi_{(2)}$ in such a way that the final equations of motion will be at most second in time derivatives and, in the process, derive the final expressions for the third and fourth order Lagrangian.

\subsubsection{Cubic Lagrangian}
$\quad$\\
\noindent Upon expanding  Eq.~(\ref{ac3}), one can verify that the terms in $\mathcal{S}_{3}$ that are expected to give rise to unwanted higher order time derivatives in the equation of motion for $\pi$ are 

\begin{equation}
\label{s3p}
\fl
\mathcal{S}_{3}^{[3]}[\pi_{(0)}^{3}]\supset\int d^{4}x a^{3} \{ \ddot{\pi}_{(0)}^{3}\left(-\tilde{M}_{1}-\tilde{M}_{5}\right)+\ddot{\pi}_{(0)}^{2}(\frac{\p_{i}^{2}\pi_{(0)}}{a^{2}}-3H\dot{\pi}_{(0)})\left(3\tilde{M}_{1}+\tilde{M}_{2}+\tilde{M}_{5}+\tilde{M}_{6}\right) \}.
\end{equation}
It is straightforward to check that the following field redefinition grants a cancellation of the above terms thanks to some of the contributions from $\mathcal{S}_{3}^{[2]}[\pi_{(0)}\pi_{(1)}]$

\bea
\label{fr1}
\fl
\label{pi1}
\pi_{(1)}=\left(\frac{\tilde{M}_{1}+\tilde{M}_{5}}{2M_{P}^{2}\dot{H}}\right)\ddot{\pi}_{(0)}^{2}+\left(\frac{3H\left(2\tilde{M}_{1}+\tilde{M}_{2}+\tilde{M}_{6}\right)}{2M_{P}^{2}\dot{H}}\right)\ddot{\pi}_{(0)}\dot{\pi}_{(0)}-\left(\frac{2\tilde{M}_{1}+\tilde{M}_{2}+\tilde{M}_{6}}{2M_{P}^{2}\dot{H}}\right)\ddot{\pi}_{(0)}\frac{\p_{i}^{2}\pi_{(0)}}{a^{2}}.\nonumber\\
\eea

\noindent The final expression for the action to third order is obtained by summing the leftover terms from the two contributions in Eq.(\ref{s3}) (while the fourth order Lagrangian is generally affected by $\pi_{(1)}$, no contributions to the cubic Lagrangian can possibly arise from $\pi_{(2)}$)

\bea
\label{s3tot}
\fl
\mathcal{S}_{3}[\pi_{(0)}]=\int d^{4}x a^{3}\{\tilde{Q}_{1}\ddot{\pi}_{(0)}\left(\frac{\p^{2}_{i}\pi_{(0)}}{a^{2}}-3H\dot{\pi}_{(0)}\right)^{2}+\tilde{Q}_{2}\left(\frac{\p^{2}_{i}\pi_{(0)}}{a^{2}}-3H\dot{\pi}_{(0)}\right)^{3}\nonumber\\\fl \qquad  +\tilde{Q}_{3}\ddot{\pi}_{(0)}\left(\frac{\p_{i}\p_{j}\pi_{(0)}}{a^{2}}-H\dot{\pi}_{(0)}\delta_{ij}\right)^{2}+\tilde{Q}_{4}\left(\frac{\p_{i}\p_{j}\pi_{(0)}}{a^{2}}-H\dot{\pi}_{(0)}\delta_{ij}\right)^{2} \left(\frac{\p^{2}_{i}\pi_{(0)}}{a^{2}}-3H\dot{\pi}_{(0)}\right)  \},\nonumber\\
\eea

\noindent where
\bea
\tilde{Q}_{1}\equiv -\tilde{M}_{1}-\tilde{M}_{2}-\tilde{M}_{3}+\tilde{M}_{6} ,\\
\tilde{Q}_{2}\equiv \tilde{M}_{1}+\tilde{M}_{2}+\tilde{M}_{3}+\tilde{M}_{4} ,\\
\tilde{Q}_{3}\equiv \tilde{M}_{5} -\tilde{M}_{7},\\
\tilde{Q}_{4}\equiv -\tilde{M}_{5}-\tilde{M}_{6}+\tilde{M}_{7}+\tilde{M}_{8}.
\eea

\noindent We can now move on to the quartic Lagrangian.

\subsubsection{Quartic Lagrangian}
$\quad$\\

\noindent The fourth order action is computed from Eq.~(\ref{s4}) with $\pi_{(1)}$ as given in Eq.~(\ref{pi1}). The expression for $\pi_{(2)}$ is to be determined in such a way that some of the contributions from  $\mathcal{S}_{4}^{[2]}[\pi_{(0)}\pi_{(2)}]$ will cancel the higher order time derivative terms from $\mathcal{S}_{4}^{[4]}[\pi_{(0)}^{4}]$, $\mathcal{S}^{[2]}_{4}[\pi_{(1)}^{2}]$ and $\mathcal{S}^{[3]}_{4}[\pi_{(0)}^{2}\pi_{(1)}]$.\\

\noindent The process of computing $\pi_{(2)}$ is straightforward but rather long, therefore we report the details in Appendix~A. Before providing the final result for the quartic Lagrangian, it is useful to make some considerations about the mass scales characterizing the action. Let us consider our initial Lagrangian (\ref{ac3})-(\ref{ac4}), with coefficients $\tilde{M}$ and $\bar{M}$, respectively for third and fourth order contributions. The field redefinition (in the form of $\pi_{(2)}$) introduces extra interactions at the quartic level that are proportional to $\hat{M}\equiv(\tilde{M}^{2}H^{2})/(M_{P}^{2}\dot{H})$. Therefore, our Lagrangian after field redefinition has the form

\bea
\mathcal{L}\sim M_{P}^{2}\dot{H}\left(\p\pi\right)^{2}+\tilde{M}\left(\partial^{2}\pi\right)^{3}+\bar{M}\left(\partial^{2}\pi\right)^{4}+\hat{M}\left(\partial^{2}\pi\right)^{4}. \label{yyy}
\eea 
Introducing the canonically normalized field $\pi_{c}\equiv M_{P}\dot{H}^{1/2}\pi$, we have 

\bea
\label{qq}
\fl
\mathcal{L}\sim \left(\p\pi_{c}\right)^{2}+\frac{\tilde{M}}{(M_{P}^{2}\dot{H})^{3/2}}\left(\p^{2}\pi_{c}\right)^{3}+\frac{\bar{M}}{M_{P}^{4}\dot{H}^{2}}\left(\p^{2}\pi_{c}\right)^{4}+\frac{\tilde{M}^{2}H^{2}}{\left(M_{P}^{2}\dot{H}\right)^{3}}\left(\p^{2}\pi_{c}\right)^{4}.
\eea
We restate the assumption (as in \cite{Creminelli1}) that all operators multiplied by a coefficient that appears at a given order for the first time, are suppressed by a common scale $\Lambda$. Eq.~(\ref{qq}) can then be rewritten as
\bea
\label{canlag}
\mathcal{L}\sim \left(\p\pi_{c}\right)^{2}+\frac{1}{\Lambda^{5}}\left(\p^{2}\pi_{c}\right)^{3}+\frac{1}{\Lambda^{8}}\left(\p^{2}\pi_{c}\right)^{4}+\frac{1}{(\tilde{\Lambda})^{8}}\left(\p^{2}\pi_{c}\right)^{4},
\eea
where we defined the scales
\bea
\label{eqr}
\Lambda^{5}\equiv\frac{M_{P}^{3}\dot{H}^{3/2}}{\tilde{M}}=\frac{M_{P}^{5/2}\dot{H}^{5/4}}{\bar{M}^{5/8}},\quad\quad\quad\quad\quad
\tilde{\Lambda}^{8}\equiv \frac{\left(M_{P}^{2}\dot{H}\right)^{3}}{\tilde{M}^{2}H^{2}}.
\eea

\noindent At this stage one can quickly relate the expression for $\bar{M}$ to the one for $\hat{M}$
\bea
\label{trueh}
\bar{M}=\hat{M}\times\left(\frac{\Lambda}{H}\right)^{2}\quad\quad\Rightarrow\quad\quad\bar{M}\gg\hat{M},
\eea
where $\Lambda\gg H$ if we are to make sense of perturbation theory. This is equivalent to stating that the scales of suppression are related by $\tilde{\Lambda}>\Lambda$. In other words, at each perturbative order, all operators multiplied by coefficients which appear for the first time at that order, are leading compared to operators that are multiplied by coefficients that appear for the first time at a lower order. Based on these considerations, we can safely neglect all operators proportional to $\hat{M}$ in the final fourth order Lagrangian which then reads: 
 
\bea
\label{final2}
\fl
\mathcal{S}_{4}=\int d^{4}x a^{3}\{\bar{Q}_{1}\frac{\left(\p_{i}^{2}\pi\right)^{4}}{a^{8}}+\bar{Q}_{2}\frac{\left(\p_{i}^{2}\pi\right)^{2}\left(\p_{j}\p_{k}\pi\right)^{2}}{a^{8}}+\bar{Q}_{3}H\dot{\pi}\frac{\left(\p_{i}^{2}\pi\right)^{3}}{a^{6}} 
+\bar{Q}_{4}H\dot{\pi}\frac{\p_{i}^{2}\pi\left(\p_{j}\p_{k}\pi\right)^{2}}{a^{6}}\nonumber\\\fl\quad\quad\quad\quad\quad\quad+\bar{Q}_{5}H^{2}\dot{\pi}^{2}\frac{\left(\p_{i}^{2}\pi\right)^{2}}{a^{4}}+\bar{Q}_{6}H^{2}\dot{\pi}^{2}\frac{\left(\p_{j}\p_{k}\pi\right)^{2}}{a^{4}}+\bar{Q}_{7}H^{3}\dot{\pi}^{3}\frac{\p_{i}^{2}\pi}{a^{2}}
\nonumber\\\fl\quad\quad\quad\quad\quad\quad+\bar{Q}_{8}\frac{\left(\p_{i}\p_{j}\pi\right)^{4}}{a^{8}}+\bar{Q}_{9}H^{4}\dot{\pi}^{4}\},
\eea

\noindent where $\bar{Q}_{1},...,\bar{Q}_{9}$ are linear combinations of the initial coefficients $\bar{M}_{i}$ (we report their exact definitions in Appendix~A). The final expression for the quartic action is now free of higher order time derivative terms and suitable for utilizing the Schwinger-Keldysh formalism \cite{Schwinger:1960qe,Calzetta:1986ey,Jordan:1986ug} in the standard Hamiltonian formulation (as in \cite{Huang:2006eha}) for the computation of the trispectrum of curvature fluctuations. Eq.~(\ref{final2}) is one of our main results.

We want to stress here that the field redefinition has not rendered the theory ghost-free, our procedure simply organizes the perturbative expansion in such a way as to have perturbations (up to a given order) to be consistent up to the scale $\Lambda$. This is, after all,  just an effective theory; if one wants to take the theory beyond this point, that is, above $\Lambda$, one should first provide arguments as to why the ghost is not there also in that regime.

\subsubsection{Interactions Hamiltonian}
$\quad$\\

\noindent At third order in the field fluctuations, the interaction Hamiltonian is given by $\mathcal{H}_{int}=-\mathcal{L}_{int}$. This relation does not generally hold true at fourth order. We quickly review the procedure for deriving $\mathcal{H}_{int}^{(4)}$ (see e.g. \cite{Huang:2006eha}, whose notation we borrow in this section). Consider a Lagrangian of the type:
\bea
\label{cl}
\fl
\mathcal{L}=f_{0}\dot{\pi}^{2}+j_{2}+ g_{0}\dot{\pi}^{3}+g_{1}\dot{\pi}^{2}+g_{2}\dot{\pi}+j_{3}+h_{0}\dot{\pi}^{4}+h_{1}\dot{\pi}^{3}+h_{2}\dot{\pi}^2+h_{3}\dot{\pi}+j_{4},
\eea
where $f_{0},\,\,g_{i},\,\,h_{j}, $ are function of $\pi$ and its spatial derivatives (the subscript indicates whether a given function is order zero ($f_{0},...$) linear ($g_{1},...$) and so on in $\pi$). Introducing the conjugate momentum
\bea
P\equiv\frac{\p \mathcal{L}}{\p \dot{\pi}},
\eea
we can derive the perturbative expansion, up to third order, of $\dot{\pi}$ in terms of powers of $P$. The total Hamiltonian reads:
\bea
\mathcal{H}(\pi,P)\equiv P\dot{\pi}-\mathcal{L},
\eea
which is split into quadratic and interactions parts
\bea
\mathcal{H}_{}=\mathcal{H}_{0}+\mathcal{H}_{int} .
\eea
In the  Schwinger-Keldysh formalism, $\mathcal{H}_{int}$ is a function of $\pi$ and $P$ as in the interaction picture ( $\pi^{I}$ and $P^{I}$). Replacing
\bea
\dot{\pi}^{I}=\frac{\p \mathcal{H}_{0}}{\p P^{I}},
\eea
in the interaction Hamiltonian, one can derive the following relations
\bea
\mathcal{H}_{int}^{(3)}=-\mathcal{L}_{int}^{(3)},\\\label{correct}
\mathcal{H}_{int}^{(4)}=-\mathcal{L}_{int}^{(4)}+\mathcal{O}\left(\frac{g_{i}g_{j}}{f_{0}}\right).
\eea
In our specific case, the corrections to $-\mathcal{L}_{int}^{(4)}$  in (\ref{correct}) are proportional to $\hat{M}=(\tilde{M}^{2}H^{2})/(M_{P}^{2}\dot{H})$. From our previous discussion on mass hierarchies, which lead to the relation $\bar{M}\gg \hat{M}$, we conclude that these corrections are subleading compared the terms in $\mathcal{L}_{int}^{(4)}$ (regulated by $\bar{M}$). In what follows we will therefore set $\mathcal{H}_{int}=-\mathcal{L}_{int}$ for the quartic order interaction Hamiltonian.\\

\section{Trispectrum of curvature fluctuations}

We now have all the ingredients necessary in order to compute the trispectrum of curvature fluctuations
\bea
\label{fd}
\langle \hat{\zeta}_{\vec{k}_{1}} \hat{\zeta}_{\vec{k}_{2}} \hat{\zeta}_{\vec{k}_{3}} \hat{\zeta}_{\vec{k}_{4}}\rangle=(2\pi)^{3}\delta^{(3)}(\vec{k}_{1}+\vec{k}_{2}+\vec{k}_{3}+\vec{k}_{4})T_{\zeta}(\vec{k}_{1},\vec{k}_{2},\vec{k}_{3},\vec{k}_{4})
\eea
where we Fourier expanded the curvature fluctuation $\zeta$:
\bea
\zeta(\vec{x},t)=\int\frac{d^{3}k}{(2\pi)^{3}}e^{i\vec{k}\cdot\vec{x}}\hat{\zeta}_{\vec{k}}(t).
\eea
Whenever metric fluctuations are neglected, as is the case here, the curvature $\zeta$ is linearly related to our field fluctuations $\pi$ \cite{Cheung}:
\bea
\zeta=-H \pi.
\eea
In particular, at leading order approximation in slow-roll, one can then essentially  read off the $\zeta$ interactions from the $\pi$ Lagrangian. Upon quantization, the curvature is expressed in terms of creation and annihilation operators
\bea
\hat{\zeta}_{\vec{k}}\equiv a_{\vec{k}}\zeta_{k}+a_{-\vec{k}}^{\dagger}\zeta_{k}^{*}.
\eea
Using conformal time variable $\tau$, it is easy to derive from the quadratic fluctuations that  the eigenfunctions are
\bea
\zeta_{k}(\tau)=\frac{-H \left(1+ik\tau\right)e^{-i k \tau}}{2\sqrt{\epsilon k^{3}}M_{P}},
\eea
where Bunch-Davies vacuum has been assumed. The power spectrum of $\zeta$ reads
\bea
\langle \hat{\zeta}_{\vec{k}}\hat{\zeta}_{\vec{k}_{'}}\rangle=(2\pi)^{3}\delta^{(3)}(\vec{k}+\vec{k}^{'})P_{\zeta}(k)\quad\quad\Rightarrow\quad\quad P_{\zeta}=\frac{H^{2}}{4M_{P}^{2}k^{3}\epsilon}.
\eea
The diagrammatic computation in the Schwinger-Keldysh formalism is outlined in Appendix~B. At tree-level, $T_{\zeta}$ is given by the sum of two distinct diagrams: a so-called \textsl{scalar-exchange} diagram, from $\mathcal{H}^{(3)}_{int}$, and a \textsl{contact-interaction} diagram from $\mathcal{H}^{(4)}_{int}$ (see Fig.~1). It turns out that the former contributions are subdominant compared to the latter. This can be easily shown by comparing the trispectrum amplitude from the two diagrams.

\noindent The amplitude of the trispectrum is conventionally described by a parameter, $\tau_{NL}$, defined as the result of the normalization of the trispectrum to the third power of the power spectrum.  $\tau_{NL}$ has a specific momentum dependence (related to the particular form of the interaction Hamiltonian) that we can only find by computing the interaction diagrams.

 However, knowing the expressions of the eigenfunctions $\zeta_{k}$ and of the interaction Hamiltonian, dimensional analysis quickly leads us to the parametric dependence of $\tau_{NL}$, which is necessary in order to estimate how big a contribution to the trispectrum a given diagram will provide. If we identify the contributions to the amplitude from the contact-interaction and from the scalar-exchange diagrams respectively as $\tau_{NL}^{c}$ and $\tau^{se}_{NL}$, we have
\bea
\tau^{c}_{NL}\sim \,\,\frac{H^{2}\bar{M}}{M_{P}^{2}\epsilon},\quad\quad\quad\quad\quad\quad\quad\quad\quad \tau^{se}_{NL}\sim \,\,\frac{H^{2}\tilde{M}^{2}}{M_{P}^{4}\epsilon^{2}},
\eea
 Let us compare $\tau^{c}_{NL}$ and $\tau^{se}_{NL}$:
\bea
\tau^{c}_{NL}\gg t^{se}_{NL}\quad\quad\quad \Leftrightarrow\quad\quad\quad \bar{M}\gg \hat{M}=\frac{\tilde{M}^2 H^2}{M_{P}^2 \dot H}.
\eea
which we have shown to be true (Eq.~(\ref{trueh})). The contribution to the total trispectrum coming from terms represented by  the contact-interaction diagram is then much larger than the contribution from the scalar-exchange side. We will therefore focus our attention on the former type of terms in the rest of the paper.

\begin{figure}[t]
	\centering
		\includegraphics[width=0.43\textwidth]{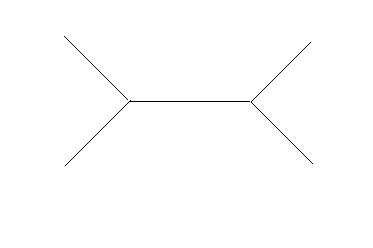}
		\hspace{8mm}
			\includegraphics[width=0.43\textwidth]{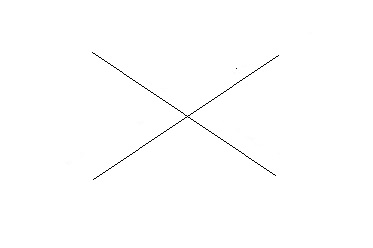}
	\label{fig:gen_dbi_limit}
		\caption{Diagrammatic representation of trispectrum contributions from $\mathcal{H}_{int}^{(3)}$ (\textsl{scalar-exchange}, left figure) and $\mathcal{H}_{int}^{(4)}$ (\textsl{contact-interaction}, right one).}
\end{figure}

\subsection{Shapes and Amplitudes from contact-interaction diagrams}

Let us rewrite the Lagrangian in Eq.~(\ref{final2}) in a schematic form

\bea
\mathcal{S}_{4}=\int d^{4}x a^{3}\left[\sum_{i=1}^{9}\bar{Q}_{i}\Theta_{i}\right],
\eea 
 where we indicated operators by $\Theta_{i}$ (e.g. $\Theta_{1}\equiv\left(\p_{i}^{2}\pi\right)^{4}/(a^{4})$ and so on). We calculated the contributions from all of the $\Theta_{i}$ interactions .  Their analytic expressions are provided in Appendix~B in the form of the shape functions $T_{\zeta}^{\Theta_{i}}$ (see Eqs.~(\ref{c1}) through (\ref{c13})). The the total trispectrum then reads:
\bea
T_{\zeta}=\sum_{i=1}^{9}T_{\zeta}^{\Theta_{i}}.
\eea
The $T_{\zeta}^{\Theta_{i}}$'s depend on the  external momenta $\vec{k}_{1}, ..., \vec{k}_{4}$ which, because of momentum conservation, are oriented so as to form a closed tetrahedron. The total number of independent variables is six. We select them to be $k_{1}$, $k_{2}$, $k_{3}$, $k_{4}$, $k_{12}$ and $k_{14}$, where $k_{i}\equiv|\vec{k}_{i}|$ and $k_{ij}\equiv|\vec{k}_{i}+\vec{k}_{j}|$. One of the $k_{i}$ may be factored out in an overall normalization factor (we are in a scale invariant case), leaving us with five variables. This is of course too large a number for a plot of the momentum dependence of the $T_{\zeta}^{\Theta_{i}}$.

One possibility is to select several different configurations in which the number of variables is narrowed down to two. We consider the configurations defined for the first time in \cite{Chen:2009bc} and applied to $P(X,\phi)$ models therein (and later often times employed by other authors, see e.g. \cite{Bartolo:2010di,Izumi:2010wm}), so as to provide a clear comparison between results for the trispectrum in our model and the corresponding outcome in other inflationary models. \\

\noindent A first glance at our results, from the qualitative appearance of the plots of the $T_{\zeta}^{\Theta_{i}}$ for each configuration, one realizes the shape functions can be classified in two main categories. The analytical significance of these two categories is related to whether the respective vertices include or not the operator $(\p_{i}\p_{j}\pi)^{2}$. It is also important to observe that the coefficients $\bar{Q}_{1}$ through $\bar{Q}_{9}$ in Eq.~(\ref{final2}) are not all independent from one another. In fact, as one can easily check by looking at their definitions in terms of the original coefficients $\bar{M}_{j}$, only four out of the nine $\bar{Q}_{i}$ are independent. One therefore expects no more than four independent contributions to $\tau_{NL}$. We select a basis of independent coefficients and collect the operators accordingly. 

\noindent The coefficients $\{\bar{Q}_{1},\bar{Q}_{2},\bar{Q}_{4},\bar{Q}_{7}\}$ are linearly independent from one another, therefore they represent a natural candidate basis for the ensemble $\bar{Q}_{1},...,\bar{Q}_{9}$. In this basis, the Lagrangian becomes

\bea
\label{anticipated}
\mathcal{S}_{4}=\int d^{4}x a^{3}\left[\bar{Q}_{1}\bar{\Theta}_{1}+\bar{Q}_{2}\bar{\Theta}_{2}+\bar{Q}_{4}\bar{\Theta}_{4}+\bar{Q}_{7}\bar{\Theta}_{7}\right],
\eea

\noindent where
\bea
\label{op1}
\bar{\Theta}_{1}\equiv \Theta_{1}-12\,\Theta_{3}+54\,\Theta_{5}-162\,\Theta_{9} ,\\
\bar{\Theta}_{2}\equiv \Theta_{2}-2\,\Theta_{3}+9\,\Theta_{5}-\frac{3}{2}\,\Theta_{8}-27\,\Theta_{9} ,\\\label{op4}
\bar{\Theta}_{4}\equiv \Theta_{4}-\Theta_{5}-\frac{3}{2}\,\Theta_{6}-\frac{1}{4}\,\Theta_{8}+\frac{9}{2}\,\Theta_{9} ,\\
\bar{\Theta}_{7}\equiv \Theta_{7}-\frac{9}{4}\,\Theta_{9} .
\eea

 \noindent From the analytic form of the diagrams with vertices $\Theta_{7}$ and $\Theta_{9}$ (Eq.~(\ref{c9})), we can immediately notice that $\bar{\Theta}_{7}$ gives a null contribution to the trispectrum. This means that we are left with three independent coefficients instead of four, which will generate three distinct contributions to $\tau_{NL}$, along with the corresponding shape functions. \noindent These shape functions are plotted in Figs.~2-3.

\noindent The expression of the total trispectrum is now:
\bea
T_{\zeta}(\vec{k}_{1},\vec{k}_{2},\vec{k}_{3},\vec{k}_{4})=T_{\zeta}^{\bar{\Theta}_{1}}+T_{\zeta}^{\bar{\Theta}_{2}}+T_{\zeta}^{\bar{\Theta}_{4}}.
\eea
The form of each of the three contributions is
\bea
T_{\zeta}^{\bar{\Theta}_{i}}=\left(\frac{H^{8}\bar{Q}_{i}}{\epsilon^{4}M_{P}^{8}}\right)\frac{\mathcal{F}_{i}}{\left(k_{1}k_{2}k_{3}k_{4}\right)^{3}}
\eea
where the $\mathcal{F}_{i}$'s are functions of momentum which can be read off from Eqs.(\ref{c1}) through (\ref{c13}), using Eqs.(\ref{op1})-(\ref{op4}). We consider the four momentum configurations introduced in \cite{Chen:2009bc}. It turns out that, in all configurations, $\mathcal{F}_{2}$ and $\mathcal{F}_{4}$ are qualitatively very similar (although their analytic expressions do not coincide), whereas $\mathcal{F}_{1}$ differs from them. As anticipated before Eq.~(\ref{anticipated}), these similarities/differences are related to the operator $(\p_{i}\p_{j}\pi)^{2}$: the vertex producing the $\mathcal{F}_{1}$ contribution, unlike the ones producing $\mathcal{F}_{2}$ and $\mathcal{F}_{4}$, does not contain $(\p_{i}\p_{j}\pi)^{2}$. We will comment on the plots of $\mathcal{F}_{1}$ and $\mathcal{F}_{2}$, in Figs.~2 and 3 respectively (all the features of $\mathcal{F}_{2}$ are shared by $\mathcal{F}_{4}$). 

\noindent We provide below a configuration-by-configuration analysis of the results.

\subsubsection*{Equilateral configuration.} 
The four external momenta have equal length ($k_{1}=k_{2}=k_{3}=k_{4}$). The leftover variables are $k_{12}$ and $k_{14}$, which we normalize to $k_{1}$. Triangular inequalities restrict their domains. If we define $x\equiv k_{12}/k_{1}$ and $y\equiv k_{14}/k_{1}$, then $x,y\in [0,2]$ and $y<\sqrt{4-x^{2}}$. As we can see from Fig.~1, $\mathcal{F}_{1}$ is constant in this configuration, a functional behaviour that is also typical of contact-interaction diagrams in $P(X,\phi)$ models \cite{Chen:2009bc}. 

On the other hand, the shape of $\mathcal{F}_{2}$ (Fig.~2) is not a plateau and it very much resembles a shape that arises from contact interaction diagrams with the operator $\left(\nabla\pi\right)^{4}$ in Ghost Inflation \cite{Izumi:2010wm}; a similar shape was found within the Effective Field Theory (EFT) approach to single-field inflation in \cite{Bartolo:2010di}, both from scalar-exchange (arising specifically from $\dot{\pi}\left(\p_{i}\p_{j}\pi\right)^{2}$ interactions) and from contact-interaction contributions (e.g. from $\left(\p_{i}\pi\right)^{4}$ and $(\p^{2}\pi)(\p_{i}\p_{j}\pi\p_{j}\p_{k}\pi\p_{k}\p_{i}\pi)$).

We stress here that the aim of the paper is manifold. One is certainly after distinct features in higher order correlators, and indeed, we do find here signatures which are not present in quite general models such as $P(X,\phi)$ models. On the other hand, we complement these findings with the fact that they occur within a theory which is, from the effective quantum field theory point of view, stable and predictive. In some respects this is in contradistinction to the approach of generic $P(X,\phi)$ models (DBI being clearly an exception) and  to the effective theory of inflation in its most comprehensive disguise \cite{Bartolo:2010di}. It is the combined (signatures \&  stability)-oriented approach which is at the heart of our work here.

\subsubsection*{Folded configuration.}
The tetrahedron has $k_{12}=0$, $k_{1}=k_{2}$ and $k_{3}=k_{4}$. The variables are $x\equiv k_{4}/k_{1}$ and $y\equiv k_{14}/k_{1}$ with domains $[0,1]$ and $[0,2]$ respectively (also $1-x\leq y \leq 1+x$). In this configuration, $\mathcal{F}_{1}$ produces a shape which is very similar to the ones we can observe from contact interactions in $P(X,\phi)$ models, whereas $\mathcal{F}_{2}$ is a slightly different version of it in such that, unlike $\mathcal{F}_{1}$, it is not constant along the $x=1$ axes. $\mathcal{F}_{2}$ also shares some features with the shapes from scalar-exchange diagrams in $P(X,\phi)$ models (also characterized by a non constant values for $x=1$) however, unlike the latter, it is not null in $(x,y)=(1,0)$ or $(1,2)$.

\subsubsection*{Specialized planar configuration.}
In this limit $k_{1}=k_{3}=k_{14}$ and $k_{12}$ can be expressed in terms of $x\equiv k_{2}/k_{1}$ and $y\equiv k_{4}/k_{1}$ (both defined in the interval $[0,2]$)
\bea
\frac{k_{12}}{k_{1}}=\sqrt{1+\frac{x^{2}y^{2}}{2}\pm\frac{xy}{2}\sqrt{\left(x^{2}-4\right)\left(y^{2}-4\right)}}.
\eea
We considered, without loss of generality, the positive sign solution for our plots. Notice that, again, $\mathcal{F}_{1}$ is very close to the contact interaction shapes for $P(X,\phi)$ models, whereas $\mathcal{F}_{2}$ is quite similar to the shape functions arising in these models from scalar-exchange diagrams.

\subsubsection*{Planar limit double-squeezed configuration.}
In this configuration $k_{3}=k_{4}=k_{12}$ and the tetrahedron is going to be flattened. This allows to write $k_{2}$ in terms of $x\equiv k_{3}/k_{1}$ and $y\equiv k_{14}/k_{1}$
\bea
\frac{k_{2}}{k_{1}}=\frac{1}{\sqrt{2}}\sqrt{1+x^{2}+y^{2}+\sqrt{3}\sqrt{-1+2x^{2}+2y^{2}+2x^{2}y^{2}-x^{4}-y^{4}}},
\eea
where $x\in [0,1]$, $y\in [0,2]$ and $1-x\leq y\leq 1+x$ apply. $\mathcal{F}_{1}$ has a shape that is very similar to the one from contact-interaction diagrams in $P(X,\phi)$ models and it becomes null as $x\rightarrow 0$; the same behaviour in the $x\rightarrow 0$ limit is observed for $\mathcal{F}_{2}$; however, the latter slighlty differs in such that, unlike the former, it goes from being positive to being begative along the $x=1$ axes, a feature which is new with respect to the existing literature.\\

\noindent To summarize our findings about shape functions, $\mathcal{F}_{1}$ noticeably reproduces the features that can be also observed in the shapes from contact-interaction diagrams in $P(X,\phi)$ models (and, thus, in DBI). 

\noindent As far as $\mathcal{F}_{2}$ is concerned, even though it arises (like $\mathcal{F}_{1}$) from contact-interaction diagrams, it presents similar features to the scalar-exchange diagrams in $P(X,\phi)$ models in the planar configuration and, but only to some extent, also in the folded configuration and in the planar limit of the double-squeezed configuration (in the latter two, $\mathcal{F}_{2}$ partially differentiates itself from the $P(X,\phi)$ results in some regions of the domain).

\noindent Intriguingly, the $\mathcal{F}_{2}$ shape in the equilateral configuration cannot be obtained at all in the $P(X,\phi)$ context and therefore represents a signature for our type of models, quite a unique one in the realm of stable theories.\\

\begin{figure}[t]
	\centering
		\includegraphics[width=0.43\textwidth]{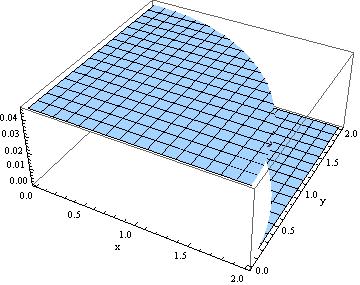}
			\includegraphics[width=0.43\textwidth]{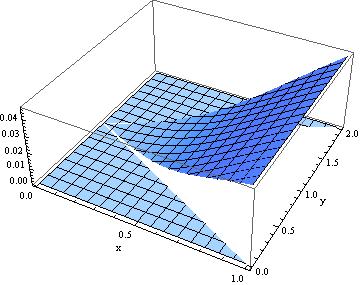}

		\includegraphics[width=0.43\textwidth]{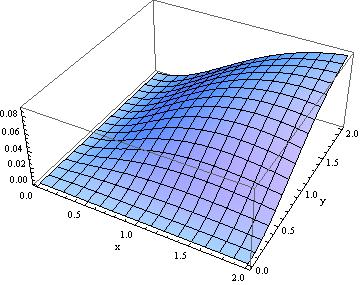}
		\hspace{8mm}
			\includegraphics[width=0.43\textwidth]{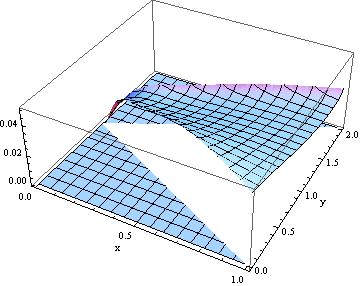}
	\caption{Plot of $\mathcal{F}_{1}$ in the equilateral (upper left), folded (upper right), specialized planar (lower left) configurations and plot of $\mathcal{F}_{1}/(k_{1}k_{2}k_{3}k_{4})$ in the planar-limit double squeezed configuration (lower right).}
	\label{ghostandgen}
\end{figure}


Let us now compute the trispectrum amplitude. Following the convention in \cite{Chen:2009bc}, we can quantify it by considering the value $\tau_{NL}$ in the limit (in \cite{Chen:2009bc} dubbed as \textsl{RT}, regular tetrahedron, limit) where $k_{1}=k_{2}=k_{3}=k_{4}=k_{12}=k_{14}\equiv k$~\footnote{Notice that $\tau_{\rm NL}$ corresponds to the $t_{\rm NL}$ quantity of~\cite{Chen:2009bc}.}
\bea
\langle \zeta^{4}\rangle\rightarrow (\pi)^{3}\mathcal{P}_{\zeta}^{3}\delta^{(3)}(\sum_{i}\vec{k}_{i})\frac{\tau_{NL}}{k^{9}}\, ,
\eea
where $\mathcal{P}_{\zeta}$ is the amplitude of the power spectrum of primordial scalar perturbations, which in our case is $\mathcal{P}_{\zeta}=(H^{2})/(\epsilon M_{P}^{2})$ . The three independent contributions to the total value of $\tau_{NL}$ read
\bea
\fl
\tau_{NL}^{(1)}=\frac{189}{512}\frac{H^{2}\bar{Q}_{1}}{\epsilon M_{P}^{2}},\quad\quad\quad \tau_{NL}^{(2)}=-\frac{4815}{8192}\frac{H^{2}\bar{Q}_{2}}{\epsilon M_{P}^{2}},\quad\quad\quad \tau_{NL}^{(4)}=-\frac{1509}{16384}\frac{H^{2}\bar{Q}_{4}}{\epsilon M_{P}^{2}}.
\eea
For detection in the CMB, $\tau_{NL}$ needs to be several orders of magnitude larger than $f_{NL}$. If we consider our Lagrangian (e.g. as in Eq.~(\ref{canlag})), we can see that 
\bea
\label{words1}
f_{NL}\zeta\sim\frac{\mathcal{L}_{3}}{\mathcal{L}_{2}}\sim \left(\frac{H}{\Lambda}\right)^{5},\quad\quad\quad\quad
\tau_{NL}\zeta^{2}\sim\frac{\mathcal{L}_{4}}{\mathcal{L}_{2}}\sim \left(\frac{H}{\Lambda}\right)^{8},
\eea
so the bispectrum is naturally easier to detect than the trispectrum. However, as the authors of \cite{Creminelli1} observed, if we compare our setup to theories with dimension six operators at the level of the cubic Lagrangian, dimension eight for the quartic Lagrangian, e.g. $P(X,\phi)$ inflation \cite{Chen:2009bc,Chen:2006nt}, we have 
\bea
\label{words2}
f_{NL}\zeta\sim\frac{\mathcal{L}_{3}}{\mathcal{L}_{2}}\sim \left(\frac{H}{\Lambda}\right)^{2},\quad\quad\quad\quad
\tau_{NL}\zeta^{2}\sim\frac{\mathcal{L}_{4}}{\mathcal{L}_{2}}\sim \left(\frac{H}{\Lambda}\right)^{4}.
\eea
Putting Eqs.~(\ref{words1}-\ref{words2}) into words: for any given value of $f_{NL}$, a larger $\tau_{NL}$ would arise in our model; precisely we have
\bea
\tau_{NL}\sim \left(\frac{f_{NL}^{4}}{\zeta}\right)^{2/5},
\eea
where we set $\zeta^{2}\sim \mathcal{P}_{\zeta}\sim 2\times 10^{-9}$.  The message is that this model is generally favored in comparison with models described by lower dimension operators (DBI being one of them), when it comes to the possibility of detecting $\tau_{NL}$.~\footnote{For constraints on some specific forms of primordial trispectra, see, e.g.~\cite{Smidt:2010ra,Fergusson:2010gn,Sekiguchi:2013hza,Ade:2013ydc}.}

As an order of magnitude estimate, for a bispectrum non-linearity parameter $|f_{\rm NL}|$ of order 1, 10, 50, and 100 one expects a $|\tau_{NL}|$ trispectrum amplitude of  the order $50$, $2 \times 10^3$, $3\times 10^{4}$, and $9 \times 10^4$, respectively.~\footnote{The values of $f_{\rm NL}$ chosen for this simple estimate are within the $95 \%$ C.L. constraints set by {\it Planck} on the equilateral and flat bispectra, which are those generated by the class of models studied here~\cite{Creminelli1}.}


\begin{figure}[t]
	\centering
		\includegraphics[width=0.43\textwidth]{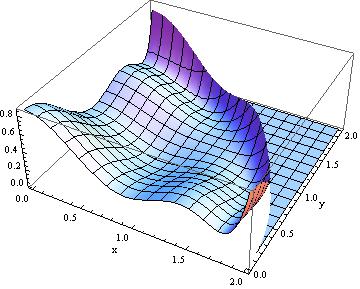}
		\hspace{8mm}
			\includegraphics[width=0.43\textwidth]{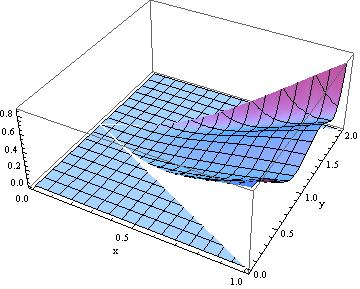}

		\includegraphics[width=0.43\textwidth]{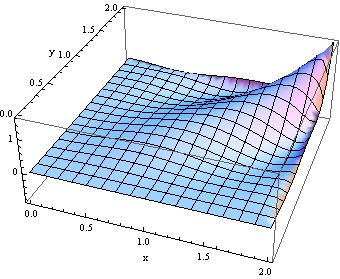}
		\hspace{8mm}
			\includegraphics[width=0.43\textwidth]{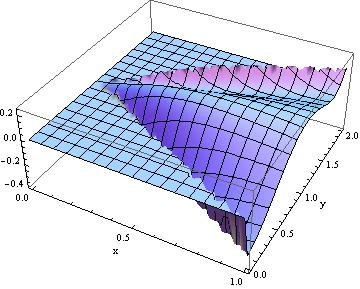}
	\caption{Plot of $\mathcal{F}_{2}$ in the equilateral (upper left), folded (upper right), specialized planar (lower left) configurations and plot of $\mathcal{F}_{2}/(k_{1}k_{2}k_{3}k_{4})$ in the planar-limit double squeezed configuration (lower right). Notably, the equilateral configuration plot is unlike any other obtained within $P(X,\phi)$ models.}
	\label{ghostandgen}
\end{figure}


\newpage
\section{Conclusions}
We have studied higher-order curvature correlators of a very general Galileon model of inflation. Despite its generality, this model retains all the essential features of Galileon models: it is stable in a quantum field theory sense and it is predictive in that interactions coefficients do not receive large corrections from renormalization. 

The key difference with respect to a seemingly analogous \cite{Andrew1, Fred} realization is that the model investigated here is better described as a ``Galileon theory of fluctuations'~\cite{Creminelli1}, as opposed to a Galileon theory of the full action. The non-renormalizations properties (with additional assumptions \cite{Creminelli1}) are expected to hold also in this model of perturbations around an FRW background.

The Lagrangian comprises more operators than that of \cite{Andrew1, Fred}, allowing for generic $\left(\nabla^2 \psi\right)^n$ interactions, of which the original Galileons of \cite{Nicolis:2008in} are a small subset. As a consequence, the equations of motion are no longer immediately second-order in time derivatives.   A field redefinition restores second order e.o.m. order by order in perturbation theory. This translates into the theory not being valid above a scale $\Lambda$ (there would be ghosts), which is a perfectly acceptable fact considering that we are seeking an effective field theory valid in a specific regime, around a specific background. 
On the other hand, having allowed for more general interactions,  one is rewarded with a richer spectrum of non-Gaussianites.

The model displays interesting non-Gaussianities already at the level of the bispectrum \cite{Creminelli1}, but, considering that other more or less stable models do the same, it is important to extend the analysis to the trispectrum so as to remove as many degeneracies as possible in the predictions. Doing so does indeed pay off: in the equilateral configuration one finds a shape-function which cannot be generated in the entire class of $P(X,\phi)$ models, of which DBI-inflation is the most notable example.

As for the impact of  {\it Planck} data  \cite{Ade:2013ydc} on primordial non-Gaussianities, the simplest single-field models of slow-roll inflation are in agreement with observations, thus passing the most stringent tests of Gaussianity performed to date. However, for what pertains alternative signatures, a window for $f_{NL}$ is still allowed, where it might be  small but not zero and might peak in configurations which are not typical of the simplest inflationary models. Also, if one is after stable models of inflation, then the number of available options shrinks considerably and the one described here is certainly one of them.
In light of the very recent results of the {\it Planck} mission \cite{Ade:2013ydc}, we stress here that the bispectrum predictions of the models studied here comprise an equilateral as well as a flat contribution, which are consistent with the constraints provided by the data~\cite{Ade:2013ydc}.

As mentioned, $\tau_{NL}$ detectability is also slightly enhanced in this model with respect to $P(X,\phi)$.  This is especially interesting if one is willing to consider  the possibility of a naturally small three-point function combined with a large four-point correlator \cite{Senatore:2010jy}. Although this does not occur in our specific model (we restrain from fine-tuning the free $\tilde{M}_n$ coefficients), it is nevertheless a feature  most likely to be realized in models highly constrained by symmetries, such as the one studied here.

As for increasing the sensitivity on measurements for non-Gaussian observables, one expects the inclusion of polarization (and more data) in {\it Planck} data analysis to improve existent constraints. The same expectations hold true for LSS, see e.g.,~\cite{Carbone:2008iz,Euclid} and possibly with CMBPol- \cite{Baumann:2008aq} and Core-like \cite{Bouchet:2011ck} future missions, as well as 21cm cosmology \cite{Cooray:2006km,Pillepich:2006fj,new21}.

\section*{Acknowledgments}

 The work of ED  was partially supported by DOE grant DE-FG02-94ER-40823 at the University of Minnesota. The work of NB has been partially supported by the ASI/INAF Agreement I/072/09/0 for the Planck LFI Activity of
Phase E2 and by the PRIN 2009 project ``La Ricerca di non-Gaussianit{\` a} Primordiale''. ED and MF are very grateful to A.J.~Tolley for many enlightening discussions. ED and MF would also like to thank the Cosmology Group at the University of Padova, and INFN, Sezione di Padova, for support and for very warm hospitality whilst parts of this work were being completed.

\section*{Appendix~A. Field redefinition}

The total action (\ref{ac4}) counts a quite large number of terms, therefore finding the right $\pi_{(2)}$ requires a long calculation. One convenient way to proceed is to work order by order in the number of temporal derivatives and cure each order at a time, starting from the highest. For the highest order, which is with ten time derivatives, we have
\bea
\fl
\mathcal{S}_{4}^{[2,t^{10}]}[\pi_{(1)}^{2}]=\int d^{4}x a^{3}\left(-4\alpha^{2}M_{P}^{2}\dot{H}\ddot{\pi}_{(0)}^{2}(\pi_{(0)}^{(3)})^{2}\right),\\
\fl
\mathcal{S}_{4}^{[3,t^{10}]}[\pi_{(0)}^{2}\pi_{(1)}]=\int d^{4}x a^{3}\left(-6\alpha\right)\left(\tilde{M}_{1}+\tilde{M}_{5}\right)\left(\ddot{\pi}_{(0)}^{2}(\pi_{(0)}^{(3)})^{2}+\ddot{\pi}_{(0)}^{3}\pi^{(4)}_{(0)}\right),\\
\fl
\mathcal{S}_{4}^{[4,t^{10}]}[\pi_{(0)}^{4}]= 0,
\eea
where $\pi_{(0)}^{(3)}\equiv d^{3} \pi_{(0)}/dt^{3}$ and $\pi_{(0)}^{(4)}\equiv d^{4} \pi_{(0)}/dt^{4}$. We used $\pi_{(1)}$ from (\ref{pi1}), which can also be rewritten as
\bea
\label{frnp}
\pi_{(1)}=\alpha\ddot{\pi}_{(0)}^{2}+3H\gamma\ddot{\pi}_{(0)}\dot{\pi}_{(0)}-\gamma\ddot{\pi}_{(0)}\left(\frac{\p_{i}^{2}\pi_{(0)}}{a^{2}}\right),
\eea
where $\alpha\equiv(\tilde{M}_{1}+\tilde{M}_{5})/2 M_{P}^{2}\dot{H}$ and $\gamma\equiv(2\tilde{M}_{1}+\tilde{M}_{2}+\tilde{M}_{6}) /2 M_{P}^{2}\dot{H}$). After integrating by parts and summing up the above contributions, the fourth order action with ten time derivatives ($\mathcal{S}_{4}^{[2]}[\pi_{(0)}\pi_{(2)}]$ excluded) reads
\bea
\fl
\mathcal{S}_{4}^{t^{10}}=\int d^{4}x a^{3} \ddot{\pi}_{(0)}^{2}(\pi_{(0)}^{(3)})^{2}\left(\frac{5(\tilde{M}_{1}+\tilde{M}_{5})^{2}}{M_{P}^{2}\dot{H}}\right).
\eea
These contributions are all unwanted and can be canceled by some of the terms $\mathcal{S}_{4}^{[2]}[\pi_{(0)}\pi_{(2)}]$ by choosing 
\bea
2M_{P}^{2}\dot{H}\pi_{(2)}\supset -\ddot{\pi}_{(0)}(\pi_{(0)}^{(3)})^{2}\left(\frac{5(\tilde{M}_{1}+\tilde{M}_{5})^{2}}{M_{P}^{2}\dot{H}}\right).
\eea
This first contribution to $\pi_{(2)}$ will also produce contributions to $\mathcal{S}_{4}^{t^{9}}$ and $\mathcal{S}_{4}^{t^{8}}$ from 
\bea
\mathcal{S}_{2}^{[2]}[\pi_{(0)}^{2}]\supset\int d^{4}x a^{3} 2  M_{P}^{2}\dot{H}\pi_{(2)}\left(3H\dot{\pi}_{(0)}-\frac{\p_{i}^{2}\pi_{(0)}}{a^{2}}\right). 
\eea
The next step will be to write down the complete expression for $\mathcal{S}_{4}^{t^{9}}$ and, similarly for what we just did for $\mathcal{S}_{4}^{t^{10}}$, find the contributions to $\pi_{(2)}$ that will cancel its unsafe terms. We can see that $\pi_{(2)}$ will receive new contributions at each order in $t^{n}$ and that, if one proceeds as for $t^{10}$, the lower order will not affect the upper ones. The total expression for $\pi_{(2)}$ then counts several contributions that can be ordered as follows
\bea
\label{frp}
\fl
\pi_{(2)}=\pi_{(2,t^{10})}+\pi_{(2,t^{9})}+\pi_{(2,t^{8})}+\pi_{(2,t^{7})}+\pi_{(2,t^{6})}+\pi_{(2,t^{5})}+\pi_{(2,t^{4})}+\pi_{(2,t^{3})},
\eea
where
\bea
\fl
2 M_{P}^{2}\dot{H}\pi_{(2,t^{10})}\equiv\ddot{\pi}_{(0)}(\pi_{(0)}^{(3)})^{2}C_{t^{10}},
\eea
\bea
\fl
2 M_{P}^{2}\dot{H}\pi_{(2,t^{9})}\equiv\dot{\pi}_{(0)}(\pi_{(0)}^{(3)})^{2}C_{t^{9}},
\eea
\bea
\fl
2 M_{P}^{2}\dot{H}\pi_{(2,t^{8})}\equiv
\ddot{\pi}_{(0)}^{3}C_{1,t^{8}}+\dot{\pi}_{(0)}^{2}\pi^{(4)}_{(0)}C_{2,t^{8}}+(\pi_{(0)}^{(3)})^{2}\frac{\p_{i}^{2}\pi_{(0)}}{a^{2}}C_{3,t^{8}} +\ddot{\pi}_{(0)}\frac{\left(\p_{i}\ddot{\pi}_{(0)}\right)^{2}}{a^{2}}C_{4,t^{8}},
\eea
\bea
\fl
2 M_{P}^{2}\dot{H}\pi_{(2,t^{7})}\equiv
\ddot{\pi}_{(0)}^{2}\dot{\pi}_{(0)}C_{1,t^{7}}+\dot{\pi}_{(0)}\pi_{(0)}^{(3)}\frac{\p_{i}^{2}\dot{\pi}_{(0)}}{a^{2}}C_{2,t^{7}}+\dot{\pi}_{(0)}\frac{(\p_{i}\ddot{\pi}_{(0)})^{2}}{a^{2}}C_{3,t^{7}}\nonumber\\\fl\qquad\qquad+\ddot{\pi}_{(0)}\frac{(\p_{i}\ddot{\pi}_{(0)})(\p_{i}\dot{\pi}_{(0)})}{a^{2}}C_{4,t^{7}}+\dot{\pi}_{(0)}\pi_{(0)}^{(4)}\frac{\p_{i}^{2}\pi_{(0)}}{a^{2}}C_{5,t^{7}},
\eea
\bea
\fl
2 M_{P}^{2}\dot{H}\pi_{(2,t^{6})}\equiv
\ddot{\pi}_{(0)}\dot{\pi}_{(0)}^{2}C_{1,t^{6}}+\dot{\pi}_{(0)}^{2}\frac{\p_{i}^{2}\ddot{\pi}_{(0)}}{a^{2}}C_{2,t^{6}}+\dot{\pi}_{(0)}\ddot{\pi}_{(0)}\frac{\p_{i}^{2}\dot\pi_{(0)}}{a^{2}}C_{3,t^{6}}+\ddot{\pi}_{(0)}^{2}\frac{\p_{i}^{2}\pi_{(0)}}{a^{2}}C_{4,t^{6}}\nonumber\\\fl\qquad\qquad+\dot{\pi}_{(0)}\frac{\p_{i}\dot{\pi}_{(0)}\p_{i}\ddot{\pi}_{(0)}}{a^{2}}C_{5,t^{6}}+\ddot{\pi}_{(0)}\left(\frac{\p_{i}^{2}\dot{\pi}_{(0)}}{a^{2}}\right)^{2}C_{6,t^{6}}+\left(\frac{\p_{i}^{2}\pi_{(0)}}{a^{2}}\right)\left(\frac{\p_{j}\ddot{\pi}_{(0)}}{a}\right)^{2}C_{7,t^{6}}\nonumber\\\fl\qquad\qquad+\ddot{\pi}_{(0)}\left(\frac{\p_{i}^{2}\pi_{(0)}}{a^{2}}\right)\left(\frac{\p_{j}^{2}\ddot{\pi}_{(0)}}{a^2}\right)C_{8,t^{6}}+\pi_{(0)}^{(4)}\left(\frac{\p_{i}^{2}\pi_{(0)}}{a^{2}}\right)^2 C_{9,t^{6}}\nonumber\\\fl\qquad\qquad+\ddot{\pi}_{(0)}\left(\frac{\p_{i}\p_{j}\ddot{\pi}_{(0)}}{a^{2}}\right)\left(\frac{\p_{i}\p_{j}\pi_{(0)}}{a^2}\right)C_{10,t^{6}}+\pi_{(0)}^{(4)}\left(\frac{\p_{i}\p_{j}\pi_{(0)}}{a^{2}}\right)^2 C_{11,t^{6}}+\ddot{\pi}_{(0)}\left(\frac{\p_{i}\p_{j}\dot{\pi}_{(0)}}{a^{2}}\right)^2 C_{12,t^{6}},\nonumber\\
\eea 
\bea
\fl
2 M_{P}^{2}\dot{H}\pi_{(2,t^{5})}\equiv \dot{\pi}_{(0)}\ddot{\pi}_{(0)}\frac{\p_{i}^{2}\pi_{(0)}}{a^{2}}C_{1,t^{5}}+\dot{\pi}_{(0)}^{2}\frac{\p_{i}^{2}\dot{\pi}_{(0)}}{a^{2}}C_{2,t^{5}}+\ddot{\pi}_{(0)}\frac{\p_{i}^{2}\pi_{(0)}}{a^{2}}\frac{\p_{j}^{2}\dot{\pi}_{(0)}}{a^{2}}C_{3,t^{5}}\nonumber\\\fl\qquad\qquad+\dot{\pi}_{(0)}\frac{\p_{i}\p_{j}^{2}\pi_{(0)}\p_{i}\ddot{\pi}_{(0)}}{a^{4}}C_{4,t^{5}}+\frac{\p_{i}\dot{\pi}_{(0)}\p_{i}\ddot{\pi}_{(0)}\p_{j}^{2}\pi_{(0)}}{a^{4}}C_{5,t^{5}}+\dot{\pi}_{(0)}\frac{\p_{i}^{2}\pi_{(0)}\p_{j}^{2}\ddot{\pi}_{(0)}}{a^{4}}C_{6,t^{5}}\nonumber\\\fl\qquad\qquad+\dot{\pi}_{(0)}\ddot{\pi}_{(0)}\frac{\p_{i}^{2}\p_{j}^{2}\pi_{(0)}}{a^{4}}C_{7,t^{5}}+\ddot{\pi}_{(0)}\frac{\p_{i}\p_{j}\pi_{(0)}\p_{i}\p_{j}\dot{\pi}_{(0)}}{a^{4}}C_{8,t^{5}}+\frac{\p_{i}\ddot{\pi}_{(0)}\p_{j}\dot{\pi}_{(0)}\p_{i}\p_{j}\pi_{(0)}}{a^{4}}C_{9,t^{5}}\nonumber\\\fl\qquad\qquad +\dot{\pi}_{(0)}\frac{\p_{i}\p_{j}\pi_{(0)}\p_{i}\p_{j}\ddot{\pi}_{(0)}}{a^{4}}C_{10,t^{5}}+\dot{\pi}_{(0)}\left(\frac{\p_{i}^{2}\dot{\pi}_{(0)}}{a^{2}}\right)^{2}C_{11,t^{5}}+\dot{\pi}_{(0)}\left(\frac{\p_{i}\p_{j}\dot{\pi}_{(0)}}{a^{2}}\right)^{2}C_{12,t^{5}},\nonumber\\
\eea
\bea
\fl
2 M_{P}^{2}\dot{H}\pi_{(2,t^{4})}\equiv  \ddot{\pi}_{(0)}\frac{\p_{i}^{2}\pi_{(0)}}{a^{2}}C_{1,t^{4}} +\ddot{\pi}_{(0)}\left(\frac{\p_{i}\p_{j}\pi_{(0)}}{a^{2}}\right)^{2}C_{2,t^{4}}+\dot{\pi}_{(0)}\frac{\p_{i}^{2}\pi\p_{j}^{2}\dot{\pi}_{(0)}}{a^{4}}C_{3,t^{4}}\nonumber\\\fl\qquad\qquad+\dot{\pi}_{(0)}\frac{\p_{i}\p_{j}\pi\p_{i}\p_{j}\dot{\pi}_{(0)}}{a^{4}}C_{4,t^{4}}+\dot{\pi}_{(0)}\frac{\p_{j}\dot{\pi}_{(0)}\p_{i}^{2}\p_{j}\pi_{(0)}}{a^{4}}C_{5,t^{4}}+\frac{\p_{i}\dot{\pi}_{(0)}\p_{j}\dot{\pi}_{(0)}\p_{i}\p_{j}\pi_{(0)}}{a^{4}}C_{6,t^{4}}\nonumber\\\fl\qquad\qquad+\frac{\left(\p_{j}\dot{\pi}_{(0)}\right)^{2}\p_{i}^{2}\pi_{(0)}}{a^{4}}C_{7,t^{4}}+\frac{\p_{i}\ddot{\pi}_{(0)}\p_{k}^{2}\pi_{(0)}\p_{i}\p_{j}^{2}\pi_{(0)}}{a^{6}}C_{8,t^{4}}+\frac{\p_{j}^{2}\ddot{\pi}_{(0)}\left(\p_{i}^{2}\pi_{(0)}\right)^{2}}{a^{6}}C_{9,t^{4}}\nonumber\\\fl\qquad\qquad+\ddot{\pi}_{(0)}\frac{\p_{i}^{2}\pi_{(0)}\p_{j}^{2}\p_{k}^{2}\pi_{(0)}}{a^{6}}C_{10,t^{4}}+\frac{\p_{i}\ddot{\pi}_{(0)}\p_{i}\p_{j}\pi_{(0)}\p_{j}\p_{k}^{2}\pi_{(0)}}{a^{6}}C_{11,t^{4}} \nonumber\\\fl\qquad\qquad           +\frac{\p_{k}^{2}\pi\p_{i}\p_{j}\ddot{\pi}_{(0)}\p_{i}\p_{j}\pi_{(0)}}{a^{6}}C_{12,t^{4}}+\frac{\p_{i}^{2}\ddot{\pi}_{(0)}\left(\p_{i}\p_{j}\pi_{(0)}\right)^{2}}{a^{6}}C_{13,t^{4}}+\frac{\left(\p_{i}^{2}\dot{\pi}_{(0)}\right)^{2}\p_{j}^{2}\pi_{(0)}}{a^{6}}C_{14,t^{4}}\nonumber\\\fl\qquad\qquad+\frac{\p_{k}^{2}\pi_{(0)}\left(\p_{i}\p_{j}\dot{\pi}_{(0)}\right)^{2}}{a^{6}}C_{15,t^{4}}+\frac{\p_{j}\ddot{\pi}_{(0)}\p_{k}\p_{m}\pi_{(0)}\p_{j}\p_{k}\p_{m}\pi_{(0)}}{a^{6}}C_{16,t^{4}} \nonumber\\\fl\qquad\qquad+\frac{\p_{k}^{2}\dot{\pi}_{(0)}\p_{i}\p_{j}\pi_{(0)}\p_{i}\p_{j}\dot{\pi}_{(0)}}{a^{6}}C_{17,t^{4}}                      ,
\eea
\bea
\fl
2 M_{P}^{2}\dot{H}\pi_{(2,t^{3})}\equiv \frac{\p_{j}^{2}\dot{\pi}_{(0)}\left(\p_{i}^{2}\pi_{(0)}\right)^{2}}{a^{6}}C_{1,t^{3}}+ \frac{\p_{i}\dot{\pi}_{(0)}\p_{k}^{2}\pi_{(0)}\p_{i}\p_{j}^{2}\pi_{(0)}}{a^{6}}C_{2,t^{3}} + \frac{\p_{k}^{2}\pi_{(0)}\p_{i}\p_{j}\pi_{(0)}\p_{i}\p_{j}\dot{\pi}_{(0)}}{a^{6}}C_{3,t^{3}} \nonumber\\\fl\qquad\qquad+ \frac{\p_{j}\dot{\pi}_{(0)}\p_{i}\p_{j}\pi_{(0)}\p_{i}\p_{k}^{2}\pi_{(0)}}{a^{6}}C_{4,t^{3}} + \frac{\p_{k}^{2}\dot{\pi}_{(0)}\left(\p_{i}\p_{j}\pi_{(0)}\right)^{2}}{a^{6}}C_{5,t^{3}} + \frac{\p_{j}\dot{\pi}_{(0)}\p_{k}\p_{i}\pi_{(0)}\p_{k}\p_{i}\p_{j}\pi_{(0)}}{a^{6}}C_{6,t^{3}},\nonumber\\
\eea
where the coefficients $C_{i,t^{n}}$ are linear combinations of $H^{n}\bar{M}_{i}$ and $H^{m}\hat{M}_{ij}$, where $\hat{M}_{ij}\equiv \tilde{M}_{i}\tilde{M}_{j}H^{2}/M_{P}^{2}\dot{H}$.\\

We have thus shown how to perform the field redefinition up to fourth order. If we neglect all operators that are proportional to $\hat{M}$ (see discussion in Sec.~2.2.2), $\pi_{(2)}$ reads

\bea
\fl
-2M_{P}^{2}\dot{H}\pi_{(2)}&=&C_{t^{8}}\ddot{\pi}_{(0)}^{3}+\left(C_{t^{7}}-3\,C_{t^{8}}\right)H\dot{\pi}_{(0)}\ddot{\pi}_{(0)}^{2}+\left(C_{1,t^{6}}+C_{t^{8}}\right)\ddot{\pi}_{(0)}^{2}\frac{\p_{i}^{2}\pi_{(0)}}{a^{2}}\nonumber\\\fl &+&\left(C_{2,t^{6}}+3\left(3C_{t^{8}}-C_{t^{7}}\right)\right)H^{2}\dot{\pi}_{(0)}^{2}\ddot{\pi}_{(0)}^{}+\Big(C_{t^{5}}-6C_{t^{8}}+C_{t^{7}}-3C_{1,t^{6}}\Big)H\dot{\pi}_{(0)}\ddot{\pi}_{(0)}^{}\frac{\p_{i}^{2}\pi_{(0)}}{a^{2}}\nonumber\\\fl&+&C_{1,t^{4}}\ddot{\pi}_{(0)}^{}\frac{\left(\p_{i}\p_{j}\pi_{(0)}\right)^{2}}{a^{4}}+\left(C_{2,t^{4}}+C_{1,t^{6}}+C_{t^{8}}\right)\ddot{\pi}_{(0)}^{}\frac{\left(\p_{i}^{2}\pi_{(0)}\right)^{2}}{a^{4}}\nonumber\\\fl&+&
C_{3,t^{4}}H^{2}\dot{\pi}_{(0)}^{2}\frac{\p_{i}^{2}\pi_{(0)}}{a^{2}}+
C_{1,t^{3}}H\dot{\pi}_{(0)}\frac{\left(\p_{i}\p_{j}\pi_{(0)}\right)^{2}}{a^{4}}+
C_{2,t^{3}}H\dot{\pi}_{(0)}\frac{\left(\p_{i}^{2}\pi_{(0)}\right)^{2}}{a^{4}}\nonumber\\\fl&+&
C_{1,t^{2}}\frac{\p_{i}^{2}\pi_{(0)}\left(\p_{i}^{2}\pi_{(0)}\right)^{2}}{a^{6}}+
C_{2,t^{2}}\frac{\left(\p_{i}^{2}\pi_{(0)}\right)^{3}}{a^{6}}.
\eea

where

\bea
\fl
C_{t^{8}}\equiv \bar{M}_{10} + \bar{M}_{11} + \bar{M}_{3} ,\\\fl
C_{t^{7}}\equiv  3 \left(2 \bar{M}_{11} + \bar{M}_{13} + 4 \bar{M}_{3} + \bar{M}_{7}\right)  ,\\\fl
C_{1,t^{6}}\equiv -\left(2 \bar{M}_{11} + \bar{M}_{13} + 4 \bar{M}_{3} + \bar{M}_{7}\right)   ,\\\fl
C_{2,t^{6}}\equiv  3  \left(-2 \bar{M}_{10} + 2 \bar{M}_{11} + 3 \bar{M}_{12} + 3 \bar{M}_{13} + \bar{M}_{14} + \bar{M}_{2} + 18 \bar{M}_{3} + 3 \bar{M}_{6} + 9 \bar{M}_{7}\right)  ,\\\fl
C_{t^{5}}\equiv 2 \left(2 \bar{M}_{10} - 2 \bar{M}_{11} - 3 \bar{M}_{12} - 3 \bar{M}_{13} - \bar{M}_{14} - \bar{M}_{2} - 18 \bar{M}_{3} - 3 \bar{M}_{6} - 9 \bar{M}_{7}\right)  ,\\\fl
C_{1,t^{4}}  \equiv  -2 \bar{M}_{10} - \bar{M}_{11} + \bar{M}_{14} + \bar{M}_{2} ,\\\fl
C_{2,t^{4}}\equiv \bar{M}_{11} + \bar{M}_{12} + \bar{M}_{13} + 6 \bar{M}_{3} + \bar{M}_{6} + 3 \bar{M}_{7},\nonumber\\\fl\\\fl
C_{3,t^{4}}\equiv -9 \left(\bar{M}_{9} - \bar{M}_{10} - \bar{M}_{11} - 3 \bar{M}_{12} - \bar{M}_{13} - \bar{M}_{14} + \bar{M}_{2} + 3 \bar{M}_{3} + 
   3 \bar{M}_{5} + 3 \bar{M}_{8} + 3 \bar{M}_{7}\right),\nonumber\\\fl\\\fl
C_{1,t^{3}}\equiv 3  \left(\bar{M}_{9} + 2 \bar{M}_{10} - \bar{M}_{11} - \bar{M}_{13} - \bar{M}_{14} + \bar{M}_{2}\right),\\\fl
C_{2,t^{3}}\equiv 2 \bar{M}_{9} - 5 \bar{M}_{10} - 2 \bar{M}_{11} - 9 \bar{M}_{12} - 2 \bar{M}_{13} - 2 \bar{M}_{14} + 2 \bar{M}_{2} + 9 \bar{M}_{3} + 9 \bar{M}_{5} + 
 9 \bar{M}_{6} + 9 \bar{M}_{7},\nonumber\\\fl\\\fl
C_{1,t^{2}}\equiv -\bar{M}_{9} - 2 \bar{M}_{10} + \bar{M}_{11} + \bar{M}_{13} + \bar{M}_{14} - \bar{M}_{2},\\\fl
C_{2,t^{2}}\equiv \bar{M}_{10} + \bar{M}_{12} - \bar{M}_{3} - \bar{M}_{5} - \bar{M}_{6} - \bar{M}_{7}.
\eea
which leads to Eq.~(\ref{final2}). The coefficients $\bar{Q}_{i}$ in Eq.~(\ref{final2}) are 

\bea
\fl
\bar{Q}_{1}&\equiv& \bar{M}_{10} + \bar{M}_{12} + \bar{M}_{4}  ,\\\fl
\bar{Q}_{2}&\equiv  &\bar{M}_{8} - 2 \bar{M}_{10} - \bar{M}_{12} + \bar{M}_{14} ,\\\fl
\bar{Q}_{3}&\equiv &-2 \bar{M}_{8} - 8 \bar{M}_{10} - 10 \bar{M}_{12} - 2 \bar{M}_{14} - 12 \bar{M}_{4} ,\\\fl
\bar{Q}_{4}&\equiv&  -4 \bar{M}_{1} - 6 \bar{M}_{8} + 8 \bar{M}_{10} + 6 \bar{M}_{12} - 2 \bar{M}_{14} ,\\\fl
\bar{Q}_{5}&\equiv&  4 \bar{M}_{1} + 15 \bar{M}_{8} + 28 \bar{M}_{10} + 39 \bar{M}_{12} + 11 \bar{M}_{14} + 54 \bar{M}_{4} ,\\\fl
\bar{Q}_{6}&\equiv&  6 \bar{M}_{1} + 9 \bar{M}_{8} - 12 \bar{M}_{10} - 9 \bar{M}_{12} + 3 \bar{M}_{14} ,\\\fl
\bar{Q}_{7}&\equiv &-12 \bar{M}_{1} - 36 \bar{M}_{8} - 9 \bar{M}_{9} - 39 \bar{M}_{10} + 9 \bar{M}_{11} - 45 \bar{M}_{12} + 9 \bar{M}_{13} - 15 \bar{M}_{14} - 
 9 \bar{M}_{2}\nonumber\\\fl&-& 27 \bar{M}_{3} - 108 \bar{M}_{4} - 27 \bar{M}_{5} - 27 \bar{M}_{6} - 27 \bar{M}_{7}  ,\\\fl
\bar{Q}_{8}&\equiv & \bar{M}_{1} + \bar{M}_{10} - \bar{M}_{14}  ,\\\fl
\bar{Q}_{9}&\equiv& 9 \bar{M}_{1} + 27 \bar{M}_{8} + \frac{81 \bar{M}_{9}}{4} + \frac{63 \bar{M}_{10}}{4} - \frac{81 \bar{M}_{11}}{4} - \frac{27 \bar{M}_{12}}{4} - \frac{81 \bar{M}_{13}}{4} - \frac{9 \bar{M}_{14}}{4} + \frac{81 \bar{M}_{2}}{4} \nonumber\\\fl&+& \frac{243 \bar{M}_{3}}{4} + 81 \bar{M}_{4} + \frac{ 243 \bar{M}_{5}}{4} + \frac{243 \bar{M}_{6}}{4} + \frac{243 \bar{M}_{7}}{4}  .
\eea

\section*{Appendix B. Computation of the Trispectrum diagrams.}

\noindent In the Schwinger-Keldysh (also dubbed as \textsl{in-in}) formalism, the general formula for the expectation value of a cosmological observable $\Theta$ at a given time $t$ is given by 
\bea
\label{sk}
\langle\hat{ \Theta}(t)  \rangle=\langle  \Big[ \bar{T}\left(e^{i\int_{0}^{t} H_{I}(t')dt'}\right)\Big]\hat{\Theta}_{I}(t) \Big[T \left(e^{-i\int_{0}^{t} H_{I}(t'')dt''}\right)  \Big] \rangle
\eea
where $T$ and $\bar{T}$ are time-ordering and anti-time ordering operators and $H_{I}$ is the interaction Hamiltonian (the subscript $I$ indicates that the fields are in the \textsl{interaction picture}). Eq.~(\ref{sk}) can equivalently be rewritten in terms of $+$ and $-$ fields
\bea
\label{sk1}
\langle \hat{\Theta}(t)  \rangle=\langle  T\left[\hat{\Theta}_{I}(t)e^{-i\int_{0}^{t} \left(H_{I}^{+}(t')-H_{I}^{-}(t')\right)dt'}\right] \rangle.
\eea
In particular, we want to compute the expectation value at late times for trispectrum of the curvature fluctuations
\bea
\fl
\langle  \hat{\zeta}_{\vec{k}_{1}}\hat{\zeta}_{\vec{k}_{2}}\hat{\zeta}_{\vec{k}_{3}}\hat{\zeta}_{\vec{k}_{4}} \rangle=(2\pi)^{3}\delta^{(3)}\left(\sum_{i=1}^{4}\vec{k}_{i}\right)T_{\zeta}\left(\vec{k}_{1},\vec{k}_{2},\vec{k}_{3},\vec{k}_{4}\right)\quad\rightarrow\quad\hat{\Theta}(t)\equiv \hat{\zeta}_{\vec{k}_{1}}\hat{\zeta}_{\vec{k}_{2}}\hat{\zeta}_{\vec{k}_{3}}\hat{\zeta}_{\vec{k}_{4}}.
\eea  
Our interaction Hamiltonian is a function of the field operator $\pi$ (and its derivatives), which is related to the curvature fluctuation by $\zeta=-H \pi$. We Fourier expand
\bea
\label{fe}
\pi(\vec{x},t)=\int \frac{d^{3}k}{(2\pi)^{3}}e^{i\vec{k}\cdot\vec{x}}\hat{\pi}_{\vec{k}}(t),
\eea
where
\bea
\hat{\pi}_{\vec{k}}(t)\equiv a_{\vec{k}}\pi_{k}(t)+a_{-\vec{k}}^{\dagger}\pi_{k}^{*}(t)
\eea
and the creation and annihilation operators obey the standard commutation relations
\bea
\Big[a_{\vec{k}},a^{\dagger}_{\vec{k}^{'}}\Big]=(2\pi)^{3}\delta^{(3)}\left(\vec{k}-\vec{k}^{'}\right).
\eea
The Feynman rules for contracting the fields read

\bea
\label{fr}
\fl
T\Big[ \pi^{+}(\vec{x},t)\pi^{+}(\vec{x}^{'},t') \Big]\quad \rightarrow\quad\langle  \pi^{}(\vec{x},t)\pi^{}(\vec{x}^{'},t')  \rangle \theta(t-t')+\langle  \pi^{}(\vec{x}^{'},t')\pi^{}(\vec{x},t)  \rangle \theta(t'-t),\\\fl
T\Big[ \pi^{+}(\vec{x},t)\pi^{-}(\vec{x}^{'},t') \Big] \quad\rightarrow\quad  \langle \pi^{}(\vec{x}^{'},t')\pi^{}(\vec{x},t)  \rangle,\\\fl
T\Big[ \pi^{-}(\vec{x},t)\pi^{+}(\vec{x}^{'},t') \Big] \quad\rightarrow\quad\langle  \pi^{}(\vec{x},t)\pi^{}(\vec{x}^{'},t')  \rangle,\\\fl
T\Big[ \pi^{-}(\vec{x},t)\pi^{-}(\vec{x}^{'},t') \Big] \quad\rightarrow\quad\langle  \pi^{}(\vec{x},t)\pi^{}(\vec{x}^{'},t')  \rangle \theta(t'-t)+\langle  \pi^{}(\vec{x}^{'},t')\pi^{}(\vec{x},t)  \rangle \theta(t-t'),
\eea
and the external lines are to be treated as $+$ fields.\\
The connected trispectrum to tree-level requires the expansion of the exponential in Eq.~(\ref{sk1}) up to second order 

\bea
\label{ue}
\fl
e^{-i\int_{0}^{t}dt^{'} \left(H_{I}^{+}(t')-H_{I}^{-}(t')\right)}\simeq &-&i\int_{0}^{t} dt^{'} \left(H_{I}^{+}(t')-H_{I}^{-}(t')\right)\\\fl&+&\frac{(-i)^{2}}{2}\int_{0}^{t}dt^{'} \left(H_{I}^{+}(t')-H_{I}^{-}(t')\right)\int_{0}^{t}dt^{''} \left(H_{I}^{+}(t'')-H_{I}^{-}(t'')\right)\nonumber.
\eea

\noindent and of the interaction Hamiltonian up to fourth order, i.e. 
\bea
H_{I}=\int d^{3}x a^{3} \left(\mathcal{H}_{int}^{(3)}+\mathcal{H}_{int}^{(4)}\right).
\eea
Two different diagrams arise from (\ref{ue}), respectively with one vertex, from $\mathcal{H}_{int}^{(4)}$, and with two vertices, from $\mathcal{H}_{int}^{(3)}$. We will refer to these as \textsl{contact-interaction} and \textsl{scalar-exchange} diagrams. The latter turn out to be subdominant compared to the former (see discussion in Sec.~3), therefore we focus on the contact-interaction ones. Their formal expression is as follows (we pick sample interaction term $\mathcal{H}_{int}^{(4)}=\bar{M}H^{4}\dot{\pi}^{4}$)

\bea
\label{ci}
\fl
T^{c.i.}_{\zeta}\left(\vec{k}_{1},\vec{k}_{2},\vec{k}_{3},\vec{k}_{4}\right)&\sim &  2H^{4} \mathcal{I}m \Big[\pi_{k_{1}}(t)\pi_{k_{2}}(t)\pi_{k_{3}}(t)\pi_{k_{4}}(t)\int_{0}^{t}dt'a^{3}\bar{M}H^{4}\dot{\pi}_{k_{1}}^{*}(t')\dot{\pi}_{k_{2}}^{*}(t')\dot{\pi}_{k_{3}}^{*}(t')\dot{\pi}_{k_{4}}^{*}(t') \Big].\nonumber\\\fl
\eea
Our eigenfunctions (in conformal time) are
\bea
\pi_{k}(\tau)=\frac{\left(1+i k \tau\right)e^{-i k \tau}}{2\sqrt{\epsilon k^{3}}M_{P}}.
\eea
The time integral in Eq.~(\ref{ci}) can be computed using the dimensionless variable $y\equiv k_{t}\tau$ (where $k_{t}\equiv k_{1}+k_{2}+k_{3}+k_{4}$) and then rescaling $y$ with a parameter $\mu$ (this will be set equal to one at the end of the calculation) and  noting that \cite{Creminelli1}
\bea
\label{pp1}
\tau e^{i\mu y}=-\frac{i}{k_{t}}\frac{d}{d\mu}e^{i\mu y},\\\label{pp2}
\tau^{2} e^{i\mu y}=-\frac{1}{k_{t}^{2}}\frac{d^{2}}{d\mu^{2}}e^{i\mu y},
\eea
and so on for higher powers of $\tau$. Note that the integrand functions are polynomials in $k \tau$ multiplied by the phase $e^{i k_{t}\tau}$. The polynomial part can be moved outside of the integral using (\ref{pp1}) and (\ref{pp2}). All we have left inside the integrals is
\bea
\int_{-\infty}^{0} dy^{}e^{i\mu y}=\frac{1}{i\mu},
\eea
which we integrated using the standard prescription of performing an infinitesimal rotation of the contour along the imaginary axis at early times. The final results for the contact interaction diagrams from Eq.~(\ref{final2}) read

\bea
\fl
\label{c1}
T_{\zeta}^{\Theta_{1}}&=&\frac{3 H^{8}\bar{Q}_{1}}{16 k_{1}k_{2}k_{3}k_{4}k_{t}^{5}\epsilon^{4}M_{P}^{8}} \Big[6+\frac{1680 k_{1}k_{2}k_{3}k_{4}}{k_{t}^{4}}+\frac{210\left(k_{2}k_{3}k_{4}+k_{1}\left(k_{3}k_{4}+k_{2}k_{3}+k_{2}k_{4}\right)\right)}{k_{t}^{3}}\nonumber\\\fl&+&\frac{30\left(k_{3}k_{4}+k_{2}\left(k_{3}+k_{4}\right)+k_{1}\left(k_{2}+k_{3}+k_{4}\right)\right)}{k_{t}^{2}}\Big]+23\,\,perms.,\\\fl
\label{c2}
T_{\zeta}^{\Theta_{2}}&=&\frac{3 H^{8}\bar{Q}_{2}\left(\hat{k}_{3}\cdot\hat{k}_{4}\right)^{2}}{16 k_{1}k_{2}k_{3}k_{4}k_{t}^{5}\epsilon^{4}M_{P}^{8}}\Big[6+\frac{1680 k_{1}k_{2}k_{3}k_{4}}{k_{t}^{4}}+\frac{210\left(k_{2}k_{3}k_{4}+k_{1}\left(k_{3}k_{4}+k_{2}k_{3}+k_{2}k_{4}\right)\right)}{k_{t}^{3}}\nonumber\\\fl&+&\frac{30\left(k_{3}k_{4}+k_{2}\left(k_{3}+k_{4}\right)+k_{1}\left(k_{2}+k_{3}+k_{4}\right)\right)}{k_{t}^{2}}\Big] +23\,\,perms. ,\\\fl
\label{c3}
T_{\zeta}^{\Theta_{3}}&=&\frac{3 H^{8}\bar{Q}_{3}}{16 k_{1}k_{2}k_{3}k_{4}k_{t}^{5}\epsilon^{4}M_{P}^{8}}\Big[1+\frac{210 k_{2}k_{3}k_{4}}{k_{t}^{3}}+\frac{30\left(k_{3}k_{4}+k_{2}\left(k_{3}+k_{4}\right)\right)}{k_{t}^{2}}+\frac{5\left(k_{2}+k_{3}+k_{4}\right)}{k_{t}}\Big]\nonumber\\\fl &+&23\,\,perms.,\\\fl
\label{c5}
T_{\zeta}^{\Theta_{4}}&=&\frac{3 H^{8}\bar{Q}_{4}\left(\hat{k}_{3}\cdot\hat{k}_{4}\right)^{2}}{16 k_{1}k_{2}k_{3}k_{4}k_{t}^{5}\epsilon^{4}M_{P}^{8}}\Big[1+\frac{210 k_{2}k_{3}k_{4}}{k_{t}^{3}}+\frac{30\left(k_{3}k_{4}+k_{2}\left(k_{3}+k_{4}\right)\right)}{k_{t}^{2}}+\frac{5\left(k_{2}+k_{3}+k_{4}\right)}{k_{t}}\Big]\nonumber\\\fl&+&23\,\,perms. ,\\\fl
T_{\zeta}^{\Theta_{5}}&=&\frac{3 H^{8}\bar{Q}_{5}}{16 k_{1}k_{2}k_{3}k_{4}k_{t}^{5}\epsilon^{4}M_{P}^{8}}\Big[1+\frac{30 k_{3}k_{4}}{k_{t}^{2}}+\frac{5\left(k_{3}+k_{4}\right)}{k_{t}}\Big]+23\,\,perms.,\\\fl
\label{c8}
T_{\zeta}^{\Theta_{6}}&=&\frac{3 H^{8}\bar{Q}_{6}\left(\hat{k}_{3}\cdot\hat{k}_{4}\right)^{2}}{16 k_{1}k_{2}k_{3}k_{4}k_{t}^{5}\epsilon^{4}M_{P}^{8}}\Big[1+\frac{30 k_{3}k_{4}}{k_{t}^{2}}+\frac{5\left(k_{3}+k_{4}\right)}{k_{t}}\Big]+23\,\,perms. ,\\\fl
\label{c9}
T_{\zeta}^{\Theta_{7}}&=&\frac{3 H^{8}\bar{Q}_{7}}{16 k_{1}k_{2}k_{3}k_{4}k_{t}^{5}\epsilon^{4}M_{P}^{8}}\Big[1+\frac{5k_{1}}{k_{t}}\Big]+23\,\,perms.=\frac{9}{4}T_{\zeta}^{\Theta_{9}},\\\fl
\label{c13}
T_{\zeta}^{\Theta_{8}}&=&\frac{3 H^{8}\bar{Q}_{8}\left(\hat{k}_{1}\cdot\hat{k}_{2}\right)^{2}\left(\hat{k}_{3}\cdot\hat{k}_{4}\right)^{2}}{16 k_{1}k_{2}k_{3}k_{4}k_{t}^{5}\epsilon^{4}M_{P}^{8}}\Big[6+\frac{1680 k_{1}k_{2}k_{3}k_{4}}{k_{t}^{4}}+\frac{210\left(k_{2}k_{3}k_{4}+k_{1}\left(k_{3}k_{4}+k_{2}k_{3}+k_{2}k_{4}\right)\right)}{k_{t}^{3}}\nonumber\\\fl&+&\frac{30\left(k_{3}k_{4}+k_{2}\left(k_{3}+k_{4}\right)+k_{1}\left(k_{2}+k_{3}+k_{4}\right)\right)}{k_{t}^{2}}\Big] +23\,\,perms. 
\eea


\end{document}